\documentclass[10pt,aps,pre,twocolumn,groupedaddress,nofootinbib,showpacs,superscriptaddress]{revtex4-1}\usepackage{graphicx}

\usepackage{color}
\definecolor{darkblue}{rgb}{0,0,0.6}
\definecolor{darkred}{rgb}{0.6,0,0}
\definecolor{darkgreen}{rgb}{0,0.6,0}

\usepackage[colorlinks=true,urlcolor=darkblue,citecolor=darkblue,linkcolor=darkred,hyperfootnotes=false]{hyperref}

\usepackage[utf8]{inputenc}
\usepackage[T1]{fontenc}
\usepackage{ae,aecompl}

\usepackage{amsmath}
\usepackage{amssymb}

\begin{document}

\title{
Finite-Time and -Size Scalings in the Evaluation of Large Deviation Functions:
\\ Numerical Approach in Continuous Time}

\author{Esteban Guevara Hidalgo}
\email[]{esteban\_guevarah@hotmail.com}
\email[\\]{nemoto@lpt.ens.fr}
\email[\\]{vivien.lecomte@univ-grenoble-alpes.fr}
\affiliation{Institut Jacques Monod, CNRS UMR 7592, Universit\'e Paris Diderot, Sorbonne Paris Cit\'e, F-750205, Paris, France}
\affiliation{Laboratoire de Probabilit\'es et Mod\`eles Al\'eatoires, Sorbonne Paris Cit\'e, UMR 7599 CNRS, Universit\'e Paris Diderot, 75013 Paris, France}

\author{Takahiro Nemoto}
\affiliation{Laboratoire de Probabilit\'es et Mod\`eles Al\'eatoires, Sorbonne Paris Cit\'e, UMR 7599 CNRS, Universit\'e Paris Diderot, 75013 Paris, France}
\affiliation{Philippe Meyer Institute for Theoretical Physics, Physics Department, \'Ecole Normale Sup\'erieure \& PSL Research University, 24 rue Lhomond, 75231 Paris Cedex 05, France}

\author{Vivien Lecomte}
\affiliation{Laboratoire de Probabilit\'es et Mod\`eles Al\'eatoires, Sorbonne Paris Cit\'e, UMR 7599 CNRS, Universit\'e Paris Diderot, 75013 Paris, France}
\affiliation{LIPhy, Universit\'e Grenoble Alpes \& CNRS, F-38042 Grenoble, France}


\begin{abstract}  
Rare trajectories of stochastic systems are important to understand -- because of their potential impact. However, their properties are by definition difficult to sample directly.
Population dynamics provides a numerical tool allowing their study, by means of simulating a large number of copies of the system, which are subjected to selection rules that favor the rare trajectories of interest.
Such algorithms are plagued by finite simulation time- and finite population size- effects that can render their use delicate.
%
%
In this paper, we present a numerical approach which uses the finite-time and finite-size scalings of estimators of the large deviation functions associated to the distribution of rare trajectories. The method we propose allows one to extract the infinite-time and infinite-size limit of these estimators which --~as shown on the contact process~-- provides a significant improvement of the large deviation functions estimators compared to the the standard one. 
%
\end{abstract}


\maketitle



\section{Introduction}

Rare events and rare trajectories can be analyzed through a variety of numerical approaches, ranging from importance sampling~\cite{kahn1951estimation}, adaptive multilevel splitting~\cite{cerou_adaptive_2007} to transition path sampling~\cite{bolhuis_transition_2002} (see \emph{e.g.}~\cite{bucklew_introduction_2013,giardina_simulating_2011} for reviews).
In this paper, we focus on population dynamics algorithms, as introduced in~\cite{giardina_direct_2006,tailleur_probing_2007}, which aims at studying rare trajectories by exponentially biasing their probability. This makes it possible to render typical the rare trajectories of the original dynamics in the simulated dynamics.
The idea is to perform the numerical simulation of a large number of copies $N_c$ of the original dynamics, supplemented with selection rules which favor the rare trajectories of interest.

The version of the population dynamics algorithm introduced by Giardinà, Kurchan and Peliti~\cite{giardina_direct_2006} provides a method to evaluate the large deviation function (LDF) associated to the distribution of a trajectory-dependent observable. The LDF is obtained as the exponential growth rate that the population would present if it was not kept constant \cite{hidalgo_discreteness_2016}.
Under this approach, the corresponding LDF estimator is in fact valid only in the limits of infinite simulation time~$t$ and infinite population size~$N_c$. The usual strategy that is followed in order to obtain those limits is to increase the simulation time and the population size until the average of the estimator over several realizations does not depend on those two parameters, up to numerical uncertainties.
The limitations and associated improvements of the population dynamics algorithm have been studied in Refs.~\cite{hurtado_current_2009,tchernookov_list-based_2010,kundu_application_2011,nemoto_population-dynamics_2016}. In this paper, following a different approach, we propose an original and simple method that takes into account the exact scalings of the finite-$t$ and finite-$N_c$ corrections in order to provide significantly better LDF estimators. 

\begin{figure}[t]
\includegraphics[width=0.48\textwidth]{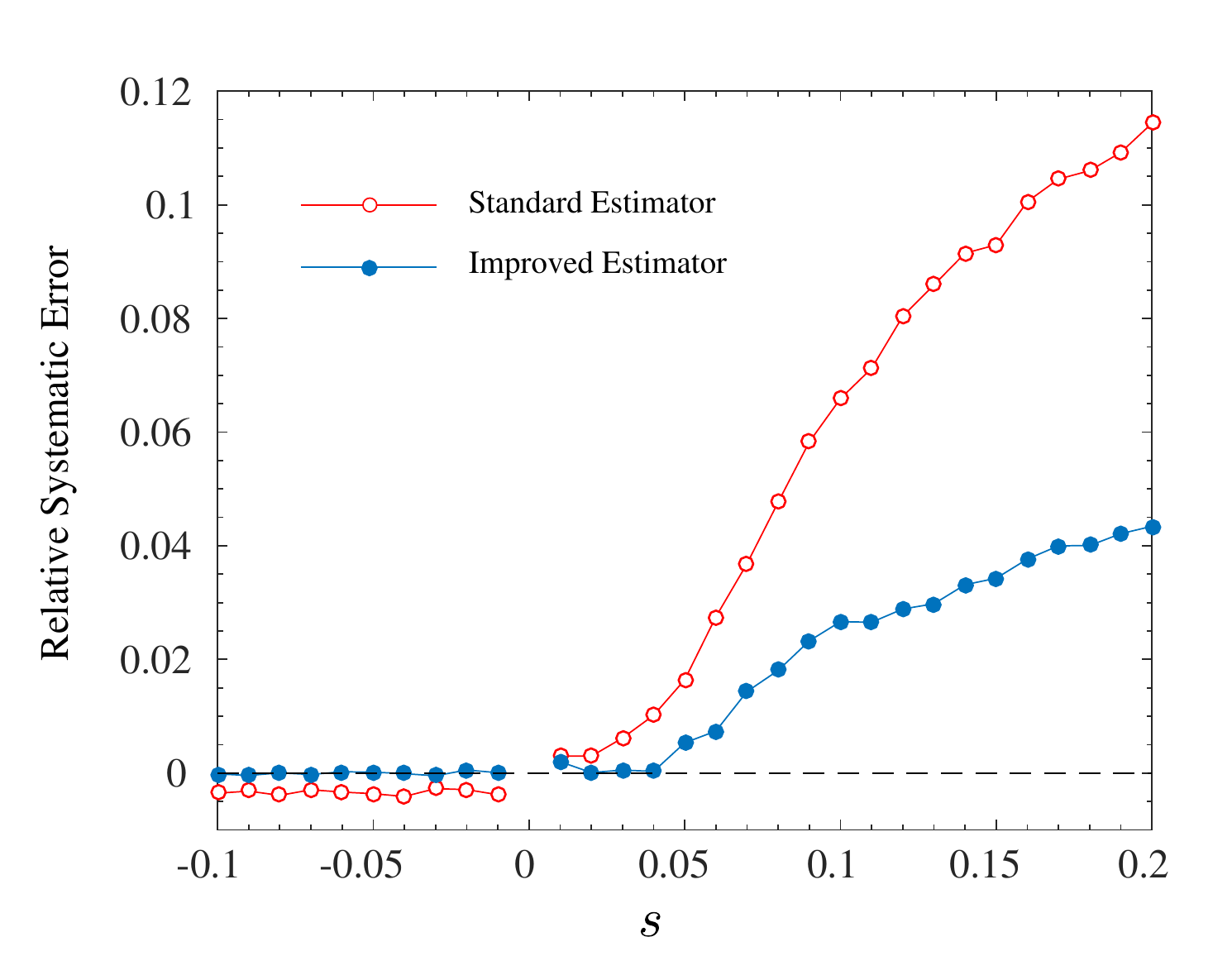}
\caption{Relative systematic error $\left [ \Psi(s)-\psi(s) \right ]/\psi(s)$ between the numerical estimators $\Psi(s)$ and the analytical LDF $\psi(s)$. The error for the standard estimator $\overline {\Psi_s^{(N_c)}(T)}$ is shown in blue and for the improved one, $f_\infty^{\infty}$ (equation~(\ref{eq:PSIinfinf})) in red. The scaling method proposed in this paper was tested on the contact process (see Sec.~\ref{subsubsec:CP}) 
(with $L = 6$, $h = 0.1$, and $\lambda=1.75$) 
for a set of populations $\vec{N}_{c} = \{20,...,200 \}$, a simulation time $T=100$, and $R =1000$ realizations. As can be seen, the errors due to finite-size and -time effects can be reduced through the improved estimator.
}

%
\label{fig:illustration-results} 
\end{figure}

In Ref.~\cite{partI}, we performed an analytical study of a {discrete-time} version of the population dynamics algorithm. We derived the finite-$N_c$ and finite-$t$ scalings of the systematic errors of the LDF estimator, showing that these behave as $1/N_c$ and $1/t$ in the large-$N_c$ and large-$t$ asymptotics respectively. 
In principle, knowing the scaling {\it a priori} means that the asymptotic limit of the estimator in the $t \to \infty$ and $N_c \to \infty$ limits may be interpolated from the data at finite $t$ and $N_c$. 
However, whether this idea is actually useful or not is a non-trivial question, as there is always a possibility that onset values of $N_c$- and $t$-scalings are too large to use these scalings. 
In the present paper, we consider a {continuous-time} version of the population dynamics algorithms~\cite{lecomte_numerical_2007,tailleur_simulation_2009}. We show numerically that one can indeed make use of these scaling properties in order to improve the estimation of LDF, in an application to a system with many-body interactions (a contact process). 
We illustrate on Fig.~\ref{fig:illustration-results} the improvement in the determination of the LDF estimator.
%
We emphasize that the two versions of the algorithm differ on a crucial point which makes that an extension of the analysis developed in~\cite{partI} cannot be done straightforwardly in order to comprehend the continuous-time case (see Appendix~\ref{Discrete-time_algorithm}). We thus stress that the observation of these scalings themselves is also non-trivial.

%
%

%
%
The paper is organized as follows. In Sec.~\ref{sec:LDF-CTalgo} we introduce the continuous-time cloning algorithm. We define the large deviations of the additive observable of interest and we detail how to estimate them. In Sec.~\ref{sec:LDFT} we study the behavior of the LDF estimator as a function of the duration of the observation time (for a fixed population~$N_{c}$) and we see how its infinite-time limit can be extracted for the numerical data.  In Sec.~\ref{sec:LDFN} we analyze the behavior of the estimator as we increase the number of clones (for a given final simulation time) and the infinite-size limit of the LDF estimator. Based on these results, we present in Sec.~\ref{sec:LDFinfTinfN} a method which allows us to extract the infinite-time, infinite-size limit of the large deviation function estimator from a finite-time, finite-size scaling analysis. Our conclusions are made in Sec.~\ref{sec:conclusion}. 
In order to complement the main discussion done through the paper we also present: In Appendix \ref{Discrete-time_algorithm}, an analysis of the difficulty of an analytical approach to the continuous-time algorithm. Then, in Appendix~\ref{sec:two-estimators}, an alternative way of defining the LDF estimator is discussed. Finally in Appendix~\ref{sec:PSIDt}, we study the fluctuations of the LDF estimator.

\section{Continuous-Time Cloning Algorithm}
\label{sec:LDF-CTalgo}
\subsection{Large Deviations of Additive Observables}
We consider a general Markov process on a discrete space of configurations $\{C\}$, with transition rates $W(C\to C')$.
The probability $P(C,t)$ for the system to be in a configuration~$C$ at time~$t$ verifies a master equation of the form $ \partial_t P = \mathbb W  P$, where the master operator $\mathbb W$ is a matrix of elements
\begin{equation}
  \label{eq:defopW}
  (\mathbb W)_{C'C}= W(C\to C') -r( C)\delta_{CC'} 
\end{equation}
and where $r(C) = \sum_{C'} W(C\to C')$ is the escape rate from configuration~$C$.
A trajectory of configurations generated in this process is denoted by $(C_0,\ldots,C_K)$, starting from $C_0$ and presenting $K$ jumps occurring at times $(t_k)_{1\leq k\leq K}$. 
We denote by $C(t')$ the state of the system at time $t'$: when $t_k \leq t' < t_{k+1}$, $C(t')=C_k$ ($k=0,1,2,\cdots, K-1$) with $t_0=0$.
We are especially interested in the large deviations of additive observables of the form
\begin{equation}
\label{eq:obs}
\mathcal O =
\sum_{k=1}^{K-1} a(C_k,C_{k+1}) + \int_0^t dt'\:b(C(t')),
\end{equation}
for trajectories of fixed duration $t$. The functions $a$ and~$b$ describe the elementary increments of the observables: $a$ accounts for quantities associated with transitions (of state), whereas $b$ does for static quantities.  
A simple example of observables of this form is that of the activity $\mathcal O = K$, which is the number of configuration changes on the time interval $[0,t]$ (in this case one has $a(C,C')=1$ and $b\equiv 0$).
We denote the joint distribution function of the state $C$ and these observables $\mathcal O$ at time $t$ by $P(C,\mathcal O,t)$.

In order to analyze large deviations of these additive observables, 
we follow the standard procedure as explained for example in \cite{lecomte_numerical_2007,tailleur_simulation_2009}. 
For this, we consider
the moment generating function
\begin{equation}
  \label{eq:MCF}
  Z(s,t)=\langle e^{-s\mathcal O}\rangle,
\end{equation}
where $\langle \cdot \rangle$ is  the expected value with respect to trajectories of duration~$t$. The parameter $s$ biases the statistical weight of histories and fixes the average value of $\mathcal O$, so that $s \neq 0$ favors its non-typical values. Since the observable $\mathcal O$ is additive and the system is described by a Markov process, 
$Z(s,t)$ satisfies at large times the scaling
\begin{equation}
  \label{eq:psi}
  Z(s,t) \sim e^{t\psi(s)}
\quad\text{for } t\to\infty,
\end{equation}
where $\psi(s)$ is the growth (or decay) rate of $Z(s,t)$ with respect to time. This exponent, known as
the scaled cumulant generating function (CGF),  is the quantity of interest in this paper. It allows ones to recover the large-time limit of the cumulants of $\mathcal O$ as derivatives of $\psi(s)$ in $s=0$, and more generically,  the distribution of $\mathcal O/t$ from the Legendre transform of $\psi(s)$~\cite{touchette_large_2009}, known as a (large deviation) rate function. Hereafter, we use the term ``large deviation function'' to refer both to the CGF and to the rate function by assuming these two are equivalent. Note that this equivalence is at least satisfied in systems that do not show any phase transition (a singularity in the rate function).

\subsection{The Mutation-Selection Mechanism}
\label{Subsec:mut}
The moment generating function $Z(s,t)$ can be computed numerically using the cloning algorithm \cite{giardina_direct_2006,tailleur_probing_2007}. In order to do that, we introduce the Laplace transform of the probability distribution $P(C,\mathcal O,t)$, defined as
\begin{equation}
  \hat P(C,s,t) = \int d\mathcal O\: e^{-s\mathcal O} P(C,\mathcal O,t).
\end{equation}
This Laplace transform allows to recover the moment generating function as $Z(s,t)=\sum_C\hat P(C,s,t)$. The probability $\hat P(C,s,t)$ satisfies a ``$s$-modified'' master equation for its time-evolution (see, \emph{e.g.}, \cite{garrahan_first-order_2009}),
\begin{equation}
 \label{eq:evolPhat}
 \partial_t \hat P = \mathbb W_s \hat P,
\end{equation}
where the ``$s$-modified'' master operator $\mathbb W_s$ is defined as
\begin{equation}
  \label{eq:defopWs}
  (\mathbb W_s)_{C'C}= W_s(C\to C') -r_s( C)\delta_{CC'} \ +\  \delta r_s( C)\delta_{CC'}.
\end{equation}
Here, $\delta r_s( C) = r_s(C)-r(C)-sb(C)$,
\begin{equation}
  \label{eq:Ws}
  W_s(C\to C') = e^{-s a(C,C')} W(C\to C')
\end{equation}
and 
\begin{equation}
  \label{eq:rs}
  r_s(C) = \sum_{C^{\prime}} W_s(C\to C'). 
\end{equation}
Contrarily to the original operator~(\ref{eq:defopW}), the ``$s$-modified'' operator~(\ref{eq:defopWs}) does not conserve probability (since $\delta r_s(C)\neq 0$), implying that there is no obvious way to simulate~(\ref{eq:evolPhat}).
However, this time-evolution equation can be interpreted not as the evolution of a single system, but as a population dynamics on a large number $N_c$ of copies of the system which evolve in a coupled way \cite{giardina_direct_2006,tailleur_probing_2007}. More precisely, reading the operator of the modified master equation (\ref{eq:evolPhat}) as in~(\ref{eq:defopWs}), we find that this evolution equation can be seen as a stochastic process of transition rates $W_s(C\to C')$ and a selection mechanism of rates
\begin{equation}
 \delta r_s( C) = r_s(C)-r(C)-sb(C).
\label{eq:deltars}
\end{equation}
where a copy of the system in configuration $C$ is copied at rate $\delta r_s( C)$ (if $\delta r_s( C)>0$) or killed at rate $|\delta r_s( C)|$ (if $\delta r_s( C)<0$).
As detailed below, the CGF $\psi(s)$ is recovered from the exponential growth (or decay) rate of a population evolving with these rules.

\subsection{Continuous-Time Population Dynamics \\ (Constant-Population Approach)}
\label{sec:cont-time-clon}
%
The mutation-selection mechanism we just described can be performed in a number of ways. One of them consists in keeping the total number of clones constant for each pre-fixed time-interval (see Refs.~\cite{giardina_direct_2006,partI} for example). 
Another one, which we use throughout this paper, consists in performing these selection mechanisms along with each evolution of the copies \cite{tailleur_simulation_2009, lecomte_numerical_2007, giardina_simulating_2011}.
A detailed description of this approach is presented below.
See also Appendix~\ref{Discrete-time_algorithm} for a brief explanation about important differences between these two techniques.

\subsubsection*{The Cloning Algorithm}
We consider $N_{c}$ clones (or copies) of the system. 
The dynamics is continuous in time: for each copy, the actual changes of configuration occur at times (which we call `evolution times') which are separated by intervals whose duration is distributed exponentially.
At a given step of the algorithm, we denote by $ \mathbf t = \{ t^{(i)} \}_{i=1,...,N_{c}}$ the set of the future evolution times of all copies and by $c = \{c_{i}\}_{i=1,...,N_{c}}$ the configurations of the copies. Their initial configurations do not affect the resulting scaled cumulant generating function in the large-time limit. However, for the concreteness of the discussion, without loss of generality, we assume that these copies have the same configuration $C$ at ${\mathbf t}=0$. The cloning algorithm is constituted of the repetition of the following procedures.
\begin{enumerate}
\item[1.] Find the clone whose next evolution time is the smallest among all the clones: Find $j={\rm argmin}_i t^{(i)}$. 
\item[2.] Compute $y_j = \lfloor Y(c_j) + \epsilon  \rfloor$, 
where the cloning factor $Y(c_j)$ is defined as $e^{\Delta t(c_j)\, \delta r_{s}(c_j)}$, $\Delta t(c_j)$ is the time spent by the clone $j$ in the configuration $c_j$ since its last configuration change, and
%
$\epsilon$ is a random number uniformly distributed on $[0,1]$.
\item[3.] If $y_j=0$, remove this copy from the ensemble, and if $y_{j}>0$, make $y_{j}-1$ new copies of this clone.
\item[4.] For each of these $y_j$ copies (if any), the state $c_{j}$ is changed independently to another state $c_{j}'$, with probability $W_{s}(c_{j} \to c_{j}')/r_{s}(c_{j})$. 
\item[5.] Choose a waiting time $\Delta t$ from an exponential law of parameter $r_{s}(c'_{j})$ for each of these copies. Its next change of configuration will occur at the evolution time $t^{(j)} + \Delta t$.
\item[6.] In order to keep the total number of copies constant, we choose randomly and uniformly: (\textit{i}) a clone $k$, $k\neq j$ and we copy it (if $y_j = 0$), or (\textit{ii}) $y_{j}-1$ clones and we erase them (if $y_j>1$).
\end{enumerate}
    
\subsection{Cumulant Generating Function Estimator}
\label{sec:cont-time-clonE}
The CGF estimator $\Psi_{s}^{(N_{c})}$ can be obtained from the algorithm we just described from the exponential growth rate that the population would present if it was not kept constant~\cite{giardina_simulating_2011}.
More precisely, this estimator is defined as
\begin{equation}
\Psi_{s}^{(N_{c})}=\frac{1}{t}\log \prod \limits_{i = 1}^{\mathcal K} X_{i},
\label{eq:PSI}
\end{equation}
where $X_{i} = (N_{c}+y_{i}-1)/N_{c}$ are the ``growth'' factors at each step $j$  of the procedure described above, and $\mathcal K$ is the total number of configuration changes in the full population up to time $t$ (which has not to be confused with $K$).
It is important to remark (as was discussed in~\cite{hidalgo_discreteness_2016} in a non-constant population context) that this growth rate can be also computed from a linear fit over the reconstructed log-population and the initial transient regime, where the discreteness effects are present, can be discarded in order to obtain a better estimation.

In practice, in order to obtain a good estimation of the CGF, it is normal to launch the simulation several times (where we denote  by $R$ the number of realizations of the same simulation), and
%
%
to estimate the arithmetic mean of the obtained values of (\ref{eq:PSI}) over these $R$ simulations. Strictly speaking (as discussed in Sec.~3.2 of~\cite{hidalgo_discreteness_2016}), as the simulation does not stop exactly at the final simulation time $T$ but at some time $t_{r}^{\mathcal{F}} \leq T$ (which is different for every $ r\in\{1,...,R\}$), the average over $R$ realizations of $\Psi_{s}^{(N_{c})}$ is then correctly defined as
\begin{equation} \label{eq:PSI1}
\overline { \Psi_{s}^{(N_{c})} }  = \frac{1}{R} \sum\limits_{r = 1}^{R}  \frac{1}{t_{r}^{\mathcal{F}}}\log\prod \limits_{i = 1}^{\mathcal K_{r}} X_{i}^{r}.
\end{equation}
However, we have observed that for not too short simulation times, $ \big \vert \overline { \Psi_{s}^{(N_{c})}(T) }  - \overline { \Psi_{s}^{(N_{c})} (t_{r}^{\mathcal{F}}) } \big \vert$ is small. By assuming $t_{r}^{\mathcal{F}} \approx T $, equation (\ref{eq:PSI1}) can be approximated by replacing $t_{r}^{\mathcal{F}}$ by $T$ (which is what we do in practice).
It is important to remark that the CGF-estimator can be defined differently from equation~(\ref{eq:PSI1}). This is done by using an alternative way of computing the average over $R$ realizations (for an example on this topic see Appendix~\ref{sec:two-estimators}). Equation~(\ref{eq:PSI1}) allows us to estimate the CGF using the constant-population approach of the continuous-time cloning algorithm for a $s$-biased Markov process, given a fixed number of clones $N_{c}$, a simulation time $T$ and $R$ realizations of the algorithm. 

\subsection{Example Models}
\label{Subsec:DefTwoState}
In order to analyze the finite-time and finite-$N_c$ scaling of the CGF estimator, we introduce two specific models: a simple two-state annihilation-creation dynamics, and a contact process on a one-dimensional periodic lattice ~\cite{lecomte_numerical_2007, CP}. In both cases, we consider the activity $K$ as the additive observable $\mathcal O$ and the analytical expression of the CGF $\psi(s)$ was obtained by solving the largest eigenvalue of the operator $\mathbb W_s$ given by~(\ref{eq:defopWs}). Below we define these models.


\subsubsection{Annihilation-Creation Dynamics}
The dynamics occurs in one site where the only two possible configurations $C$ are either $0$ or $1$. The transition rates are
\begin{equation} \label{eq:W_AC}
 W(0 \rightarrow 1) = c \text{    ,    } W(1 \rightarrow 0) = 1-c,
\end{equation}
where $c\in [0,1]$.
%
%
%
The analytical expression for the CGF of the activity in this case corresponds to
\begin{equation} \label{eq:PSIA}
\psi({s})=-\frac{1}{2}+\frac{1}{2} \bigg( 1-4c(1-c)(1-e^{-2s}) \bigg)^{1/2}.
\end{equation}

\subsubsection{Contact Process}
\label{subsubsec:CP}
Each position $i$ of a $L$-sites one-dimensional lattice is occupied by a spin which is either in the state $n_{i}=0$ or $n_{i}=1$. The configuration $C$ is then constituted by the states of these spins, {\it i.e.}, $C=(n_i)_{i=1}^L$. The dynamics occurs on this lattice with periodic boundary conditions 
with transition rates 
$W(n_{i}=1 \rightarrow n_{i}=0) = 1$ and 
\begin{equation} \label{eq:W_CP}
W(n_{i} = 0 \rightarrow n_{i} = 1) = \lambda(n_{i-1}+n_{i+1}) + h,
\end{equation}
where $\lambda$ and $h$ are positive constants.
This model is an example of contact processes \cite{CP}, which have been studied in many contexts especially for the spread of infections \cite{10.2307/2244329}. It has been known that the corresponding CGF develops a singularity in $L \rightarrow \infty$, showing a dynamical phase transition~\cite{lecomte_numerical_2007, Lecomte2007}.

\section{Finite-time and Finite-$N_c$ Behavior of CGF Estimator}

In this section, we focus on the annihilation-creation process for a peculiar value of $s$ ($s=-0.2$), which is representative of the full range of $s$ on which we study large deviations. 

\subsection{Finite-Time Scaling}
\label{sec:LDFT}

\begin{figure} [ht]
\includegraphics[width=0.49\textwidth]{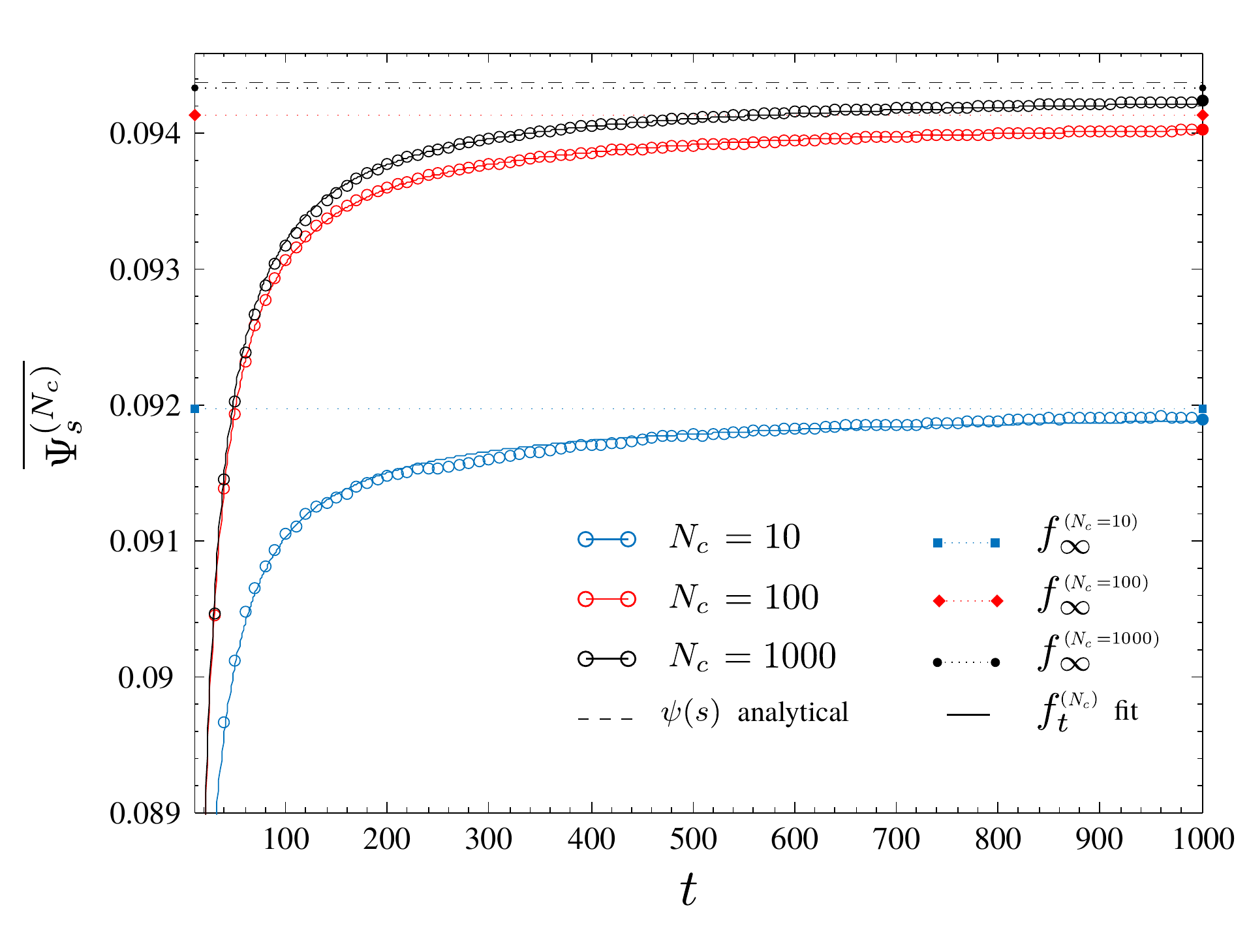}
\caption{\label{fig:averagePsi_1} 
Average over  $R = 10^{4}$ realizations of the CGF estimator $\Psi_{s}^{(N_{c})}$ (equation~(\ref{eq:PSI1})) as a function of duration $t$ of the observation window, for $N_{c} \in \{10,100,1000 \}$ clones, for the annihilation-creation dynamics~(\ref{eq:W_AC}) with $c=0.3$. The analytical expression for the large deviation function $\psi(s)$ (equation~(\ref{eq:PSIA})) is shown with a black dashed line and the fitting functions $f_{t}^{(N_{c})}$ encoding the finite-$t$ scaling (equation~(\ref{eq:fitdef})) are shown with continuous curves. The (\emph{a priori}) best estimation of the large deviation function (to which we refer as standard estimator) is given by $\overline{ \Psi_{s}^{(N_{c})}(t)}$ at the largest simulation time $T = 1000$, which are shown with solid circles (at the right end of the figure). The extracted infinite-time limits $f_{\infty}^{(N_{c})}$ are shown as dotted lines and squares ($N_{c} = 10$), diamonds ($N_{c} = 100$) and circles ($N_{c} = 1000$). }
\end{figure}

Here, we study the large-time behavior of the CGF estimator, at fixed number of clones $N_c$. 
Fig.~\ref{fig:averagePsi_1} presents the average over $R=10^4$ realizations of the CGF estimator $\overline{ \Psi_{s} ^{(N_{c})} }$ as a function of the (simulation) time for given numbers of clones $N_{c} = \{10,100,1000 \}$. It is compared with the analytical value $\psi(s)$ (equation~(\ref{eq:PSIA})) which is shown with a black dashed line. 
%

As can be seen in Fig.~\ref{fig:averagePsi_1} for a small number of clones ($N_{c}=10$),
the CGF estimator $\overline{ \Psi_{s} ^{(N_{c})} }$ is highly deviated from the analytical value $\psi(s)$. However, as $N_c$ and the simulation time $t$ become larger, the CGF estimator get closer to the analytical value $\psi(s)$.

One can expect that in the $t \to \infty$ and $N_c \to \infty$ limits, $\psi(s)$ will be obtained from the estimator as
\begin{equation}
\lim_{N_c \to \infty} \lim_{t \to \infty} \overline{\Psi_{s}^{(N_{c})}(t)} = \psi(s),
\label{eq:limitNcT}
\end{equation}
as it was derived in~\cite{partI}. However, in a practical implementation of the algorithm, this infinite-time and -size limits are not achievable and we use large but {\it finite} simulation time $t$ and number of clones $N_{c}$. 
This fact motivates our analysis of the actual dependence of the estimator with $t$ and $N_{c}$.
The standard estimator of the large deviation function is the value of $\overline{ \Psi_{s} ^{(N_{c})} }$ at the largest simulation time $T$ and for the largest number of clones $N_c$, (\emph{e.g.}, $\overline{ \Psi_{s} ^{(N_{c})}(T) }$ for $N_{c}=1000$ and $T=1000$, the black solid circle~{\small$\bullet$} in Fig.~\ref{fig:averagePsi_1}). This value provides the (\emph{a priori}) best estimation of the large deviation function that we can obtain from the continuous-time cloning algorithm. 
However encouragingly, as we detail later, this estimation can be improved by taking into account the convergence speed of the CGF estimator. 

%

The result of fitting $\overline{ \Psi_{s} ^{(N_{c})}(t)}$ with the curve $f_{t}^{(N_{c})}$ defined as
\begin{equation}
f_{t}^{(N_{c})}  \equiv f_{\infty}^{(N_{c})} + b_{t}^{(N_{c})}t^{-1}
\label{eq:fitdef}
\end{equation}
is shown with solid lines in Fig.~\ref{fig:averagePsi_1}. The fitting parameters $f_{\infty}^{(N_{c})}$ and $b_{t}^{(N_{c})}$ can be determined from the least squares method by minimizing the deviation from $\overline{ \Psi_{s} ^{(N_{c})}(t)}$. 
The clear coincidence between $\overline{ \Psi_{s} ^{(N_{c})}(t)}$ and the fitting lines indicates the existence of a $1/t$-convergence of $\overline{ \Psi_{s} ^{(N_{c})}(t)}$ to $\lim_{t\rightarrow \infty}\overline{ \Psi_{s} ^{(N_{c})}(t)}$ (that we call $1/t$-scaling). 
%
This property can be derived from the assumption that the cloning algorithm itself is described by a Markov process: in \cite{partI} with a different version of the algorithm, we constructed a meta-Markov process to describe the cloning algorithm 
by expressing the number of clones by a birth-death process. 
Once such meta process is constructed, the CGF estimator (\ref{eq:PSI}) is regarded as the time-average of the observable $X_i$ within such meta-Markov process%
\footnote{
In other words, $t \Psi_{s}^{(N_{c})}$ is an additive observable of the meta-process describing the cloning algorithm, as read from~(\ref{eq:PSI}).
}%
.%
We now recall that time-averaged quantities converge to their infinite-time limit with an error proportional to $1/t$ when the distribution function of the variable converges exponentially (as in Markov processes). This leads to the $1/t$-scaling of CGF estimator (\ref{eq:fitdef}). 
%
We note that constructing such meta-Markov process explicitly is not a trivial task, and for the algorithm discussed here, such a construction 
remains as an open problem. 

By assuming the validity of the scaling form~(\ref{eq:fitdef}), it is possible to extract the infinite-time limit of the CGF estimator from finite-time simulations. We denote this infinite-time limit as $f_{\infty}^{(N_{c})}$ and it is expected to be a 
the better estimator of CGF than $\overline{ \Psi_s^{(N_c)}(T)}$ at finite $T$, provided that
\begin{equation}
f_\infty^{(N_c)} = \lim _{t\rightarrow \infty }\overline{ \Psi_s^{(N_c)}(t)}.
\label{eq:infT}
\end{equation}
%
In Fig.~\ref{fig:averagePsi_1},
we show $f_{\infty}^{(N_{c})}$ with dotted lines and circles ($N_{c} = 10$), diamonds ($N_{c} = 100$) and squares ($N_{c} = 1000$).
As can be seen, 
this parameter indeed provides a better numerical estimate of $\psi({s})$ than $\overline{ \Psi_{s} ^{(N_{c})}(T)}$. 

\subsection{Finite-$N_c$ Scaling \label{sec:LDFN}}
\label{sec:LDFT_Nc}

\begin{figure}[t]
\includegraphics[width=0.48\textwidth]{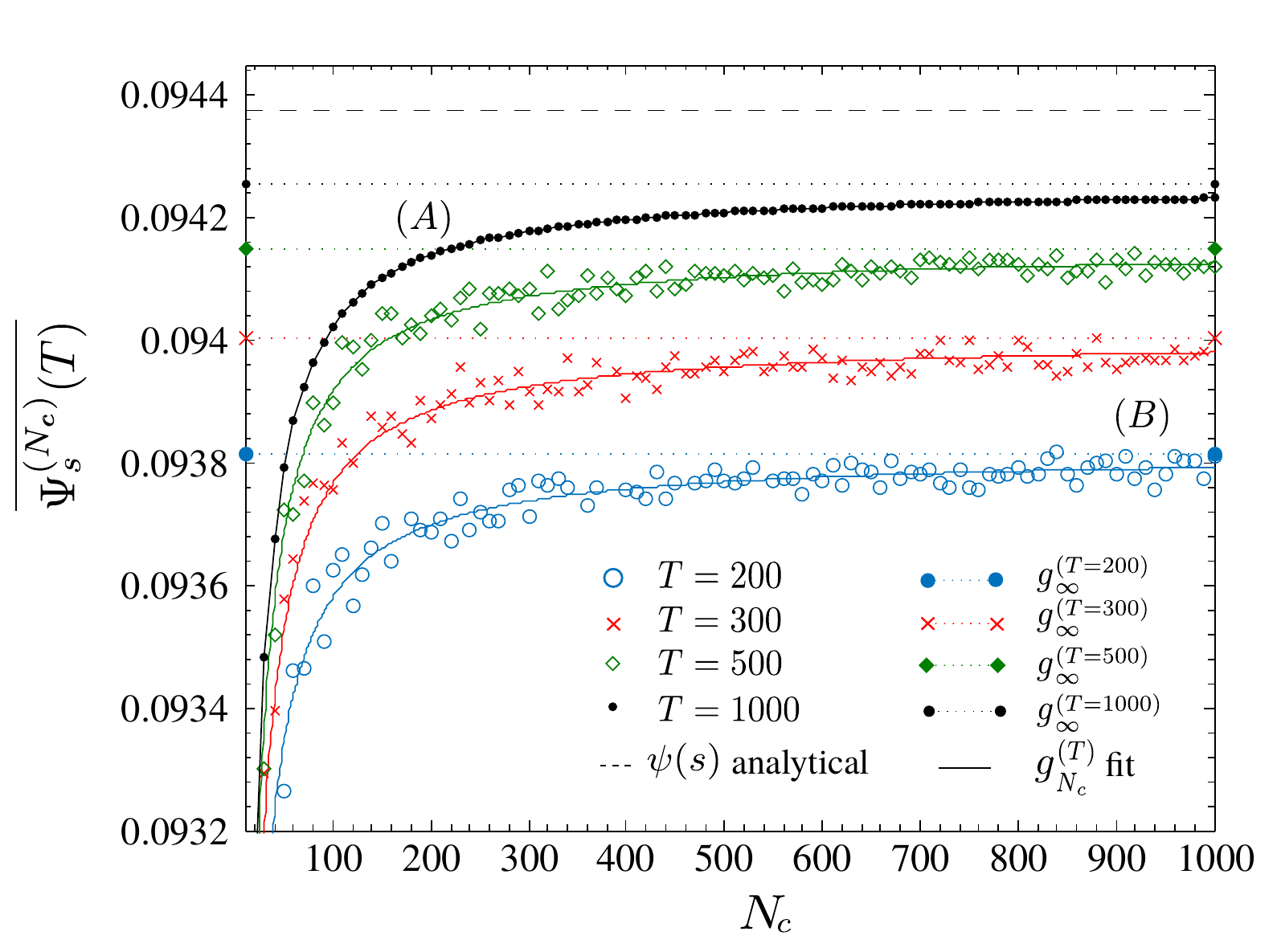}
\caption{\label{fig:PsiN} CGF Estimator $\overline{ \Psi_{s}^{(N_{c})} (T) }$ (equation~(\ref{eq:PSI1})) for given final (simulation) times $T = \{200,300,500, 1000 \}$ as a function of the number of clones $N_c$ (on the range $10\leq N_{c}\leq 1000$). The analytical value $\psi(s)$~(\ref{eq:PSIA}) is shown with a dashed line and the fits $g_{N_c}^{(T)}$ (equation~(\ref{eq:PSI4N})) with continuous curves. A large simulation time for a small number of clones, shown in~(\emph{A}),  produces a better estimation compared to the one given by the largest number of clones with a relatively short simulation time, which is shown in~(\emph{B}). The best CGF estimation we can naively obtain would be given by $\overline{ \Psi_{s}^{(N_{c})}(T) }$ at largest simulation time $T$ and largest number of clones $N_{c}$. However, the extracted infinite-size limits $g_{\infty}^{(T)}$ provide a better estimation in comparison. These limits are shown with dotted lines and circles ($T=200$), crosses ($T=300$), diamonds ($T=500$) and dots ($T=1000$). Additionally, $c=0.3$ and $s=-0.2$.
} 
\end{figure}

Here, we study the behavior of the CGF estimator $\overline{ \Psi_{s}^{(N_{c})} (T) }$ as we increase the number of clones $N_c$, for a given final (simulation) time $T$. 
%
%
%
Similar to what we did in Sec.~\ref{sec:LDFT}, we consider a curve in the form
\begin{equation} \label{eq:PSI4N}
g_{N_{c}}^{(T)} = g_{\infty}^{(T)} + \tilde b_{N_{c}}^{(T)}N_{c}^{-1},
\end{equation}
where $g_{\infty}^{(T)}$ and $\tilde b_{N_{c}}^{(T)}$ are fitting parameters which are determined by the least squares fitting to 
$\overline{ \Psi_{s}^{(N_{c})} (T) }$. The obtained $g_{N_{c}}^{(T)}$ as a function of $N_{c}$ are shown in Fig.~\ref{fig:PsiN} as solid lines.   
We considered four values of final simulation time $T = \{200,300,500, 1000 \}$ and population sizes in the range $10\leq N_{c}\leq 1000$. 
As can be seen, these curves describe well the dependence in $N_c$ of $\overline{ \Psi_{s}^{(N_{c})} (T) }$, indicating that $\overline{ \Psi_{s}^{(N_{c})} (T) }$ converges to its infinite-$N_c$ limit with an error proportional to $1/N_c$ (that we call $1/N_c$-scaling). 
This scaling could be proved under general assumptions in~\cite{partI}, ($i$) however 
%
without covering the continuous-time algorithm discussed here,
and ($ii$) for the CGF estimator $\overline{ \Psi_{s}^{(N_{c})} (T)}$ considered the $T \to \infty$ limit, instead of finite $T$. The generalization of the argument presented in~\cite{partI} in order to cover the general cases ($i$) and ($ii$) is an important open direction of  research.

By assuming the validity of such $1/N_c$-scaling,
we can evaluate
the $N_c \to \infty$ limit of $\overline{ \Psi_{s}^{(N_{c})} (T) }$ as
the fitting parameter $g_{\infty}^{(T)}$ obtained from finite $N_c$ simulations as
\begin{equation}
g_{\rm \infty}^{(T)} =  \lim_{N_c\to \infty}  \overline{ \Psi_{s}^{(N_{c})} (T) }.
\end{equation} 
These parameters $g_{\infty}^{(T)}$ (to which we refer as infinite-size limit) are shown in Fig.~\ref{fig:PsiN} as dotted lines.  As shown in the figure, $g_{\rm \infty}^{(T)}$ provides
better estimations of $\psi({s})$ than the one given by the standard estimator $\overline{ \Psi_{s}^{(N_{c})} (T) }$. 
%


Complementary to the discussion done in this section, in Appendix \ref{sec:PSIDt} we analyze the fluctuations of the CGF estimator.

\section{Finite-Time and Finite-$N_c$ Scaling Method to estimate Large Deviation Functions  \label{sec:LDFinfTinfN}}

In the previous section, we have shown how
it is possible to extract $f^{(N_c)}_\infty$ and $g^{(T)}_\infty$ from finite $T$- and finite $N_c$- simulations respectively. In this section, we combine both of these $1/t$- and $1/N_c$- scaling methods in order to extract the infinite-time and -size limit of the CGF estimator. This limit gives a better evaluation of the large deviation function within the cloning algorithm than the standard estimator. 


\begin{figure} [!t]
\includegraphics[width=0.48\textwidth]{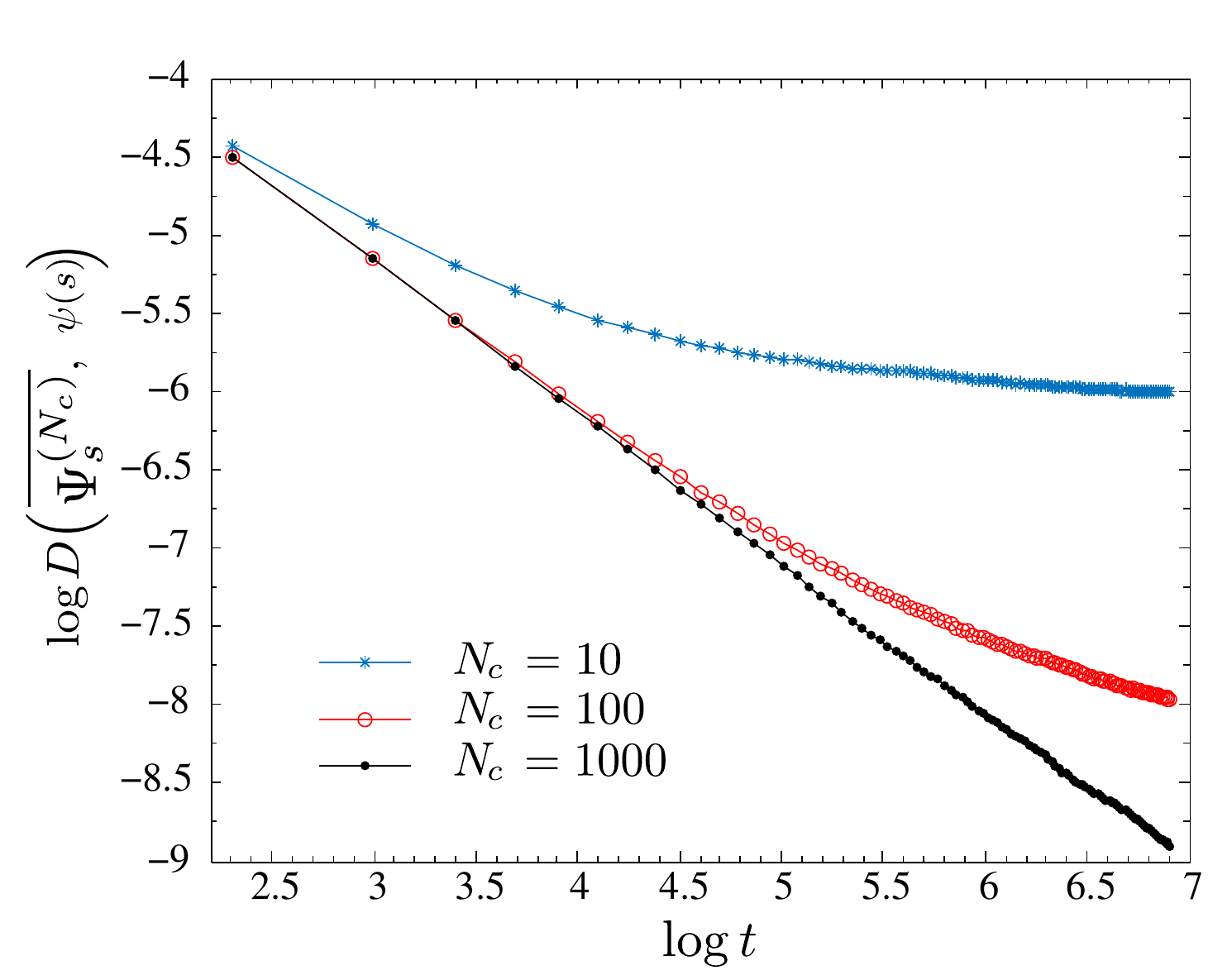}
\caption{\label{fig:averagePsi_2} 
Distance $D$ (equation~\eqref{eq:D1}) between the analytical CGF $\psi(s)$ and its numerical estimator $\overline{ \Psi_{s}^{(N_{c})} }$, as a function of time~$t$ in log-log scale. The distances are computed from the values in Fig.~\ref{fig:averagePsi_1}. This distance behaves as a power law of exponent $-1$ on a time window, where the size of the time window increases as $N_{c}$ increases. This illustrates the scaling~\eqref{eq:D2}. The parameters of the model are $c = 0.3$, $s=-0.2$.}
\end{figure}

We first note that either of $f_{\infty}^{(N_c)}$ or $g_{\rm \infty}^{(T)}$
%
is expected to converge to $\psi(s)$ as $N_c \to \infty$ or as $T \to \infty$. 
We checked numerically this property by defining the distance $D$ between $\psi({s})$ and its numerical estimator $\overline{ \Psi_{s}^{(N_{c})}}$, 
\begin{equation} \label{eq:D1}
D \big(  \overline{\Psi_{s}^{(N_{c})} },\psi(s) \big) = \big \vert \overline{ \Psi_{s}^{(N_{c})} } - \psi(s) \big \vert.
\end{equation}
This quantity is shown in Fig.~\ref{fig:averagePsi_2} as a function of $t$ in log-log scale. As we can see, as $N_{c}$ increases, $\log D$ behaves as straight line with slope $-1$ on a time window which grows with $N_c$. In other words, when $N_{c}\to\infty$, 
\begin{equation} \label{eq:D2}
\big \vert \overline{ \Psi_{s}^{(N_{c})} } - \psi(s) \big \vert \sim t^{-1}.   
\end{equation}


Inspired by this observation, we assume the following scaling
for the fitting parameter $f_{\infty}^{N_{c}}$. If we consider a set of simulations performed at population sizes $\vec{N}_{c}= \{ N_{c}^{(1)},...,N_{c}^{(j)} \}$, the obtained infinite-time limit of the CGF estimator $f_{\infty}^{N_{c}}$ behaves as a function of $N_{c}$ as
\begin{equation} \label{eq:PSIinfinf}
 f_{\infty}^{(N_c)} \simeq f_{\infty}^{\infty} + b_{\infty}^{(N_c)} N_c^{-1},
\end{equation}
which means that $f_{\infty}^{(N_c)}$ itself exhibits $1/N_c$ corrections for large but finite $N_c$. 
By using this scaling, we detail below in Sec.~\ref{ssec:SM} the method to extract the infinite-time infinite-$N_c$ limit of the CGF estimator $\overline{\Psi_{s}^{(N_{c})}(T)}$ from finite-time and finite-$N_c$ data. We note that this method can be used for a relatively short simulation time and a relatively small number of clones (see Fig.~\ref{fig:PSIT-PSIN}). In Sec.~\ref{ssec:exCP}, we 
present numerical examples of the application of this method to the contact process. 

\subsection{The Scaling Method}
\label{ssec:SM}
The procedure is summarized as follows:
\begin{enumerate}
\item[1.] Determine the average over $R$ realizations $\overline{ \Psi_{s}^{(N_{c})}(t) }$ (equation~\eqref{eq:PSI1}) up to a final simulation time $T$ for each $N_c \in \vec{N}_{c}$.
\item[2.] Determine the fitting parameter $f_{\infty}^{(N_{c})}$ defined in the form $f_{t}^{(N_{c})} = f_{\infty}^{(N_{c})} + b_{t}^{(N_{c})}t^{-1}$ from each of the obtained $\overline{ \Psi_{s}^{(N_{c})}(t) }$.  
\item[3.] Determine $f_{\infty}^{\infty}$ from a fit in size $f_{\infty}^{(N_c)} = f_{\infty}^{\infty} + b_{\infty}^{(N_c)}N_c^{-1}$ (equation~\eqref{eq:PSIinfinf}) on $f_{\infty}^{(N_{c})}$.
\end{enumerate}

The result obtained for $f_{\infty}^{\infty}$ renders a better estimation of $\psi(s)$ than the standard estimator $\overline{ \Psi_{s}^{(N_{c})}(t) }$ evaluated for $N_c = \max \vec{N_c}$ and for $t = T$. 

%


\subsection{Application to the Contact Process}
\label{ssec:exCP}
We apply the scaling method to the one-dimensional contact process (see Sec.~\ref{Subsec:DefTwoState} for the definition). 
We set $L = 6$, $h = 0.1$, $\lambda=1.75$, $T=100$ and $s=0.15$. 
%
%
As we detail below, we compare the improved estimator $f_{\infty}^{\infty}$ obtained from 
the application of the scaling method (for $\vec{N}_{c} = \{20,40,...,180,200 \}$) with the standard estimator $\overline{ \Psi_{s}^{(N_{c})} (T) }$ (for $N_{c}=\max \vec{N}_{c} = 200$).
\begin{figure}[ht]
\includegraphics[width=0.48\textwidth]{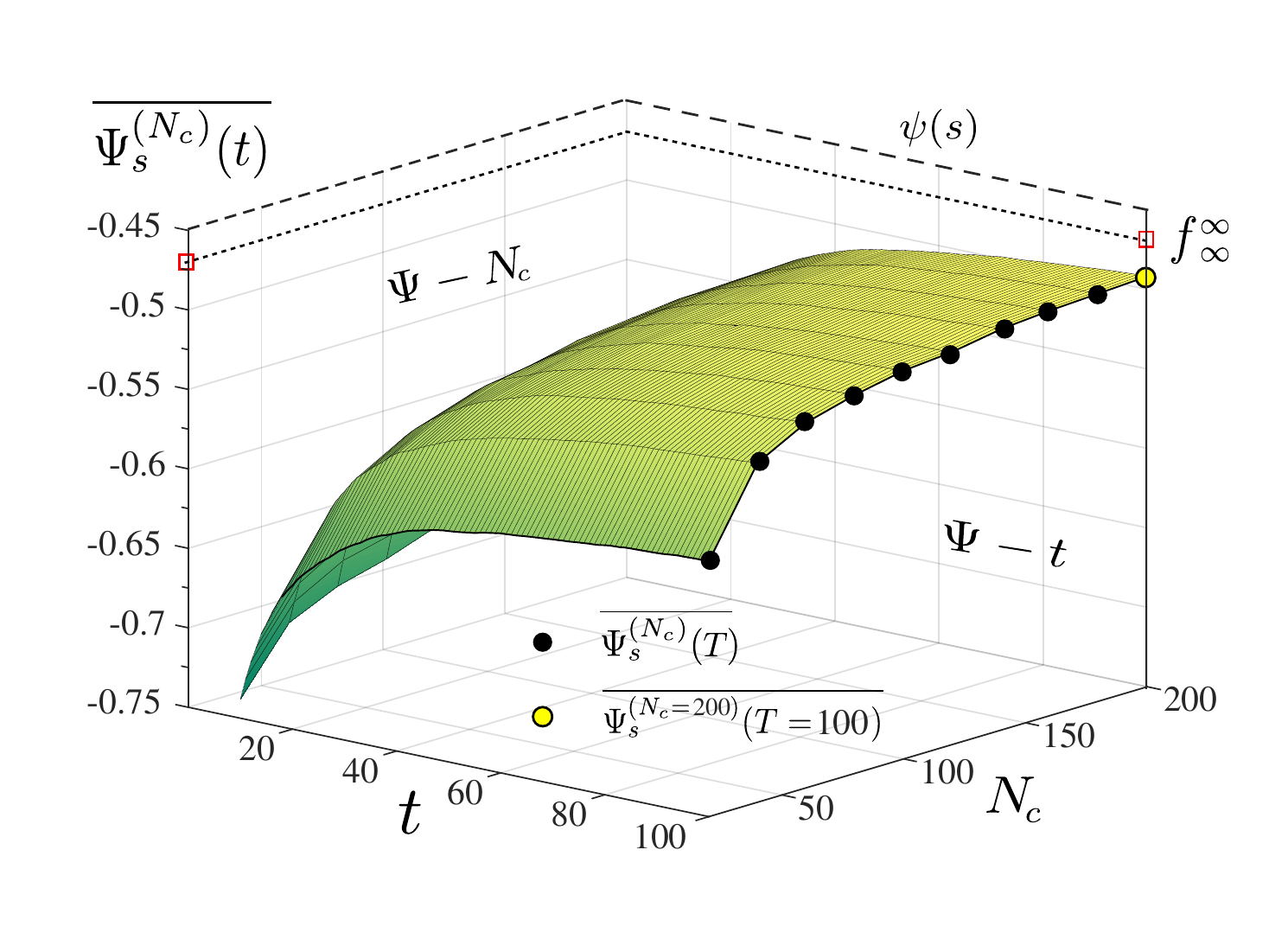}
\caption{\label{fig:surfPSI} Estimator of the large deviation function $\overline{ \Psi_{s}^{(N_{c})} (t) }$ as a function of time and the number of clones. The estimator $\overline{ \Psi_{s}^{(N_{c})} (T) }$ at final simulation time $T = 100$ as a function of the number of clones (up to $N_{c}=200$) is shown as black circles. The best CGF estimation under this configuration given by the standard estimator, \emph{i.e.},  $\overline{ \Psi_{s}^{(N_{c}=200)} (T=100) }$ is shown as a yellow circle. The analytical value of the CGF  $\psi({s})$ is obtained from the largest eigenvalue of the operator~(\ref{eq:defopWs}) and shown as a black dashed line. The extracted limit $f_\infty^\infty$ is shown with red squares. Additionally, $L=6$, $s=0.15$, $h = 0.1$, $\lambda=1.75$ and $R=10^{3}$.
} 

\end{figure}
\begin{figure*} [t]
\includegraphics[width=0.48\textwidth]{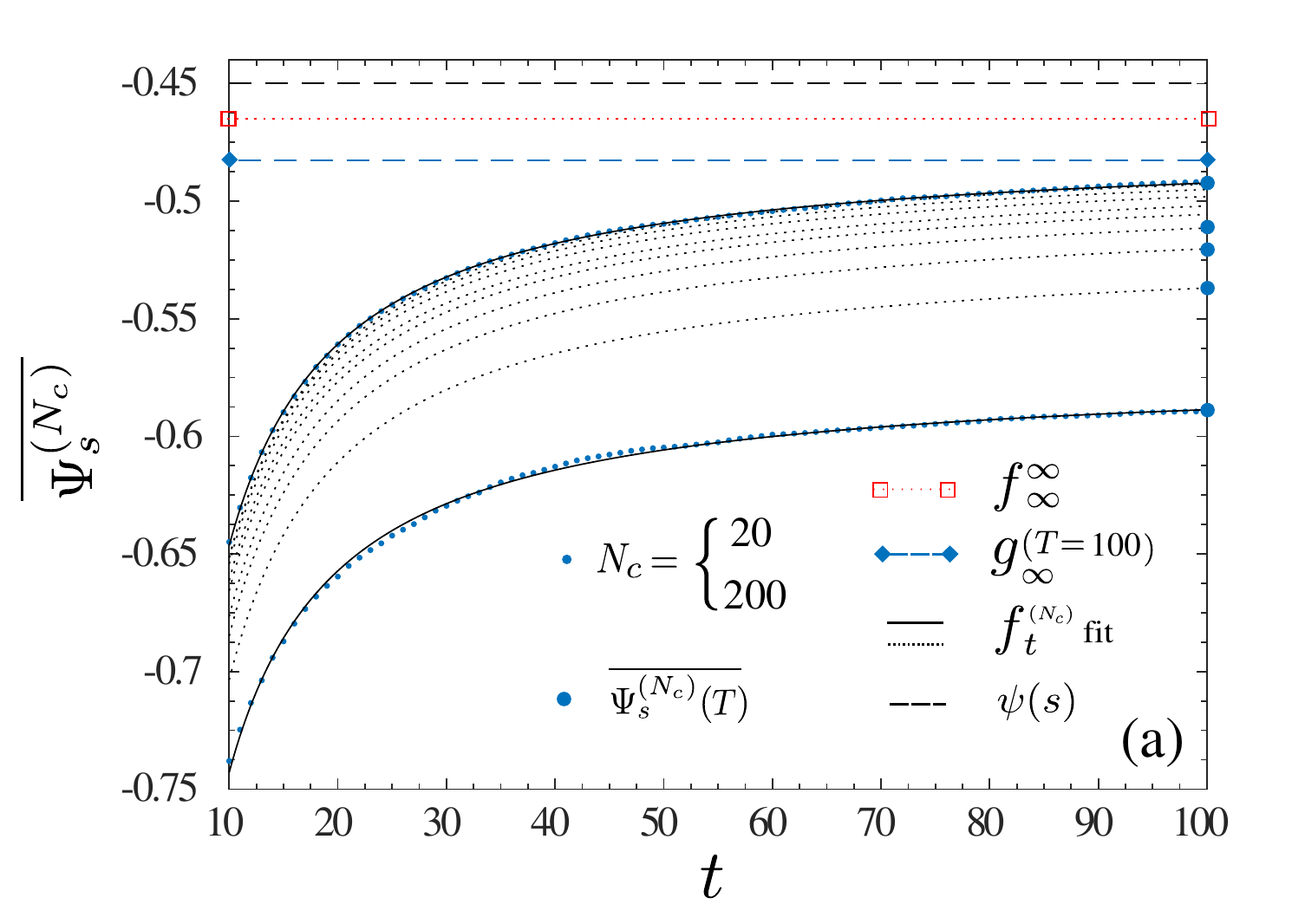}
\includegraphics[width=0.48\textwidth]{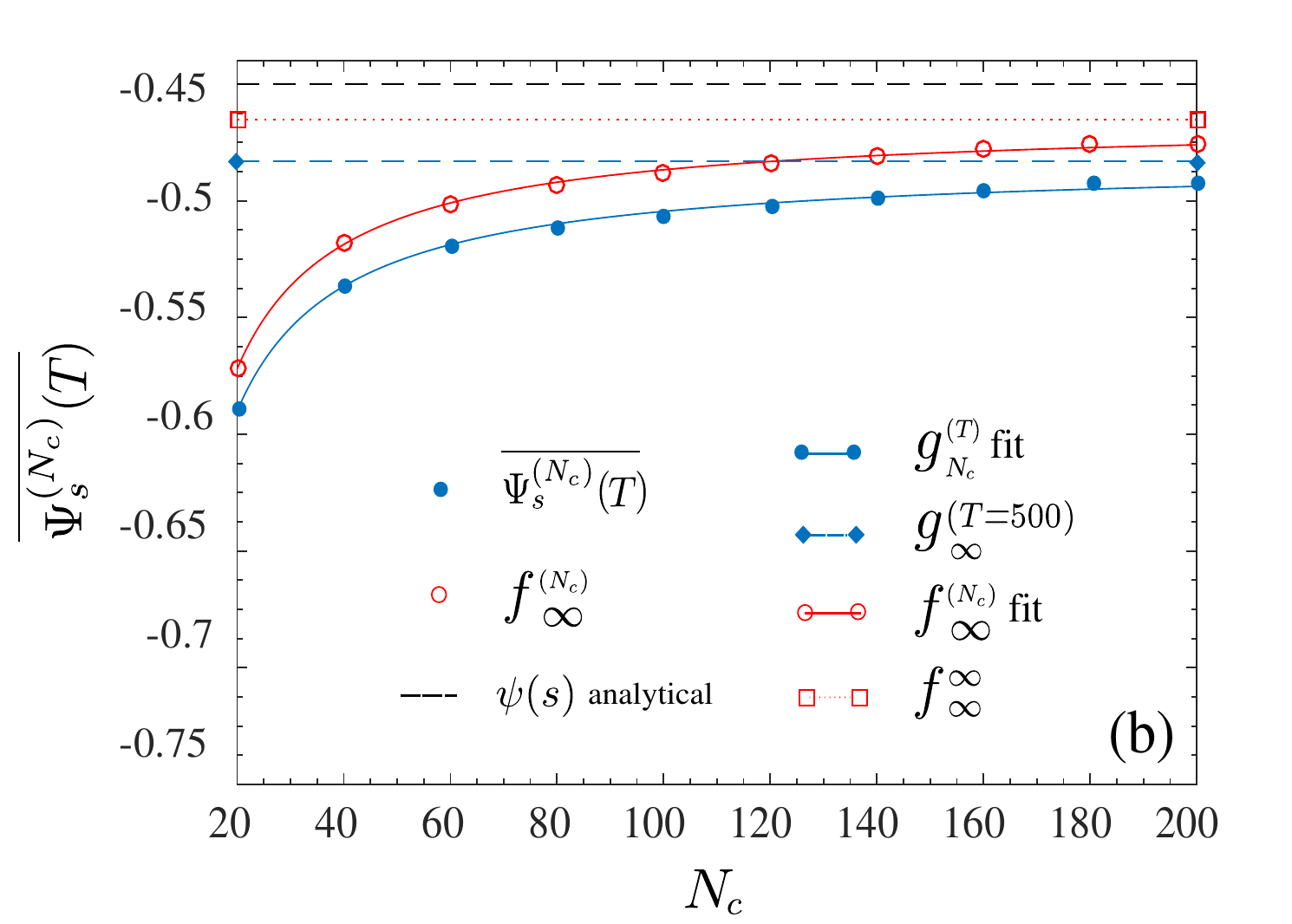}
\caption{\label{fig:PSIT-PSIN} \textbf{(a)} Projection of the surface represented in Fig.~\ref{fig:surfPSI} over the plane $\Psi-t$. $\overline{ \Psi_{s}^{(N_{c})}(t) }$ is represented for $N_{c} = 20$ and $N_{c}= 200$ with blue dots. The estimations $\overline{ \Psi_{s}^{(N_{c})}(T) }$ of the large deviation (at the final simulation time $T=100$) are shown in large blue dots for all the values of $N_{c}$ considered. 
The fit in time (equation~\eqref{eq:fitdef}) over $\overline{ \Psi_{s}^{(N_{c})}(t)}$ is shown as black solid lines (for $N_{c} = 20$ and $N_{c}= 200$) and dotted lines (for other values of $N_c$). 
\textbf{(b)} Projection at the final simulation time $T=100$ on the plane $\Psi-N_{c}$, $\overline{ \Psi_{s}^{(N_{c})}(T) }$ is shown in large blue dots.
The infinite-time limit $f_{\infty}^{(N_{c})}$ as a function of $N_c$ (see equation (\ref{eq:fitdef})) is represented in red circles.
The results of fitting $\overline{ \Psi_{s}^{(N_{c})}(T) }$ (equation~(\ref{eq:PSI4N})) and  $f_{\infty}^{(N_{c})}$ (equation~(\ref{eq:PSIinfinf})) are shown with blue and red solid curves respectively. The infinite-$N_c$ limit $g_{\infty}^{(T)}$ is shown with blue dashed line and diamonds meanwhile the infinite-size and time limit $f_{\infty}^{\infty}$ is shown with a red dotted line in both of \textbf{(a)} and \textbf{(b)}. 
The extracted limit $f_{\infty}^{\infty}$ renders a better estimation of the large deviation function than $\overline { \Psi_{s}^{(N_{c} = 200)}(T = 100) }$ (and also than $g_{\infty}^{(T)}$) demonstrating the efficacy of the method proposed.
}
\end{figure*}

Fig.~\ref{fig:surfPSI} represents the behavior of the estimator $\overline{ \Psi_{s}^{(N_{c})}(t) }$ as a function of the simulation time~$t$ and of the number of clones $N_{c}$. 
%
%
%
The values of the estimator at the final simulation time~$T$ are represented with black circles for each $N_{c} \in \vec{N}_{c}$ and with a yellow circle for $N_{c} = \max \vec{N}_{c}$. 
The analytical expression for the large deviation function $\psi({s})$ is shown in a black dashed line.


On Fig.~\ref{fig:PSIT-PSIN}(a) we show the projection of the surface of Fig.~\ref{fig:surfPSI} on the plane $\Psi-t$. The behavior in~$t$ of the estimator $\overline{ \Psi_{s}^{(N_{c})}(t) }$ is shown for $N_{c} = 20$ and $N_{c} = 200$, in blue dots in Fig.~\ref{fig:PSIT-PSIN}(a). The standard CGF estimators, $\overline{ \Psi_{s}^{(N_{c})}(T) }$, are shown in large blue dots in Fig.~\ref{fig:PSIT-PSIN}(a) (on the axis for $T=100$).
The fitting curves $f_{t}^{(N_{c})}$ (equation~(\ref{eq:fitdef})) are shown in black continuous lines (for $N_c=20$ and $N_c=200$) and black dotted lines (for other intermediate values of $N_c$). 
Next, we show in Fig.~\ref{fig:PSIT-PSIN}(b) the projection of the surface of Fig.~\ref{fig:surfPSI} on the plane $\Psi-N_c$ where the time has been set to the largest $t=T$. 
The standard CGF estimators, $\overline{ \Psi_{s}^{(N_{c})}(T)}$ are plotted as
blue filled circles, and the fitting curve $g_{N_c}^{(T)}$ (equation (\ref{eq:PSI4N})) on $\overline{ \Psi_{s}^{(N_{c})}(T)}$ is shown as a blue solid line. From these curves, we determine  $g_{\infty}^{(T)}$ (see Sec.~\ref{sec:LDFT_Nc}), which is shown as a blue dashed line and diamonds. 
Finally, the parameter $f_{\infty}^{(N_{c})}$ extracted from the fitting on $\overline{ \Psi_{s}^{(N_{c})}(t)}$ (for each value of $N_{c}$) is shown as red circles in Fig.~\ref{fig:PSIT-PSIN}(b). These values also scale as $1/N_c$ (equation (\ref{eq:PSIinfinf})) and their fit is shown as a red solid curve. 
The scaling parameter $f_{\infty}^{\infty}$ obtained from this last step  provides a better estimation of the large deviation function than the standard estimator $\overline { \Psi_{s}^{(N_{c} = 200)}(T = 100) }$ that is widely used in the application of cloning algorithms. 
%
%
This improvement is valid on a wide range of values of the parameter $s$ as can be visualized in Fig.~\ref{fig:illustration-results}, where we represented the relative systematic error $\left [ \Psi(s)-\psi(s) \right ]/\psi(s)$ between the standard and improved estimators $\Psi(s)$ and the analytical LDF~$\psi(s)$.

\section{Conclusion}
\label{sec:conclusion}

Direct sampling of the distribution of rare trajectories is a rather difficult numerical issue (see for instance~\cite{rohwer_convergence_2015}) because of the scarcity of the non-typical trajectories. We have shown how to increase the efficiency of a commonly used numerical method (the so-called cloning algorithm) in order to improve the evaluation of large deviation functions which quantify the distribution of such rare trajectories, in the large time limit.
We used the finite-size and finite-time scaling behavior of CGF estimators in order to propose an improved version of the continuous-time cloning algorithm which provides more reliable results, 
%
less affected by finite-time and -size effects.
We verified the results observed for the discrete-time version of the cloning algorithm~\cite{partI} and we showed their validity also for the continuous case. Importantly, we showed how these results can be applied to more complex systems.

%

We note that the scalings which rule the convergence to the infinite-size infinite-time limits (with corrections in $1/N_c$ and  in $1/t$) have to be taken into account properly: indeed, as power laws, they present no characteristic size and time above which the corrections would be negligible.
The situation is very similar to the study of the critical depinning force in driven random manifolds: the critical force presents a corrections in one over the system size~\cite{kolton_uniqueness_2013} which has to be considered properly in order to extract its actual value.
Generically, such scalings also provide a convergence criterion to the asymptotic regimes of the algorithm: one has to confirm that the CGF estimator does present corrections (first) in $1/t$ and (second) in $1/N_c$ with respect to an asymptotic value in order to ensure that such value does represent a correct evaluation of the LDF.

It would be interesting to extend our study of these scalings to systems presenting dynamical phase transitions (in the form of a non-analyticity of the CGF), where it is known that the finite-time and the finite-size scalings of the CGF estimator can be very hard to overcome~\cite{lecomte_numerical_2007}. 
In particular, in this context, it would be useful to understand how the dynamical phase transition of the original system translates into anomalous features of the distribution of the CGF estimator in the cloning algorithm. 
These phase transitions are normally accompanied with an infinite system-size  limit (although there was a report of dynamical phase transitions without taking a such limit \cite{0295-5075-116-5-50009}). To overcome these difficulties (caused by a large system size and/or by the presence of a phase transition), it may be useful to use the adaptive version of the cloning algorithm \cite{PhysRevLett.118.115702}, which has been recently developed to study such phase transitions, with the scaling method presented in this paper.

\begin{acknowledgements}
 E.~G.~thanks Khashayar Pakdaman for his support and discussions. Special thanks to the Ecuadorian Government and the Secretar\'ia Nacional de Educaci\'on Superior, Ciencia, Tecnolog\'ia e Innovaci\'on, SENESCYT. T.~N.~gratefully acknowledges the  support of Fondation Sciences Math\'ematiques de Paris -- EOTP NEMOT15RPO, PEPS LABS and LAABS Inphyniti CNRS project.
%
V.~L~acknowledges support by the ERC Starting Grant 680275 MALIG and by the ANR-15-CE40-0020-03 Grant LSD.
\end{acknowledgements}

\appendix

\section{Issues on an Analytical Approach \label{Discrete-time_algorithm}}

In a previous analytical study~\cite{partI}, we considered a \emph{discrete-time} version of the population dynamics algorithm, where a cloning procedure is performed every small time interval $\Delta t$. 
We have proved the convergence of the algorithm in the large-$N_c$, -$t$ limits,
and we also derived that the systematic error of the LDF estimator (\emph{i.e.}, the deviation of the estimator from the desired LDF) decayed proportionally to $1/N_c$ and $1/t$. 
From a practical point of view, however, the formulation used there had one problem. In order to prove the result, we took the large frequency limit of cloning procedure or, in other words, we took the $\Delta t \rightarrow 0$ limit. 
A rough estimate of the error due to non-infinitesimal $\Delta t$ proves to be $O(\Delta t)$. For a faster algorithm, it is better to take this value to be larger, and indeed empirically, we expect that this error to be very small (or rather disappearing in the large $t, N_c$ limits). However, the detailed analytical estimation of this error 
is still an open problem. 

In the main part of this current manuscript, from a different point of view,  we consider the \emph{continuous-time} version of the population dynamics algorithm 
%
%
\cite{lecomte_numerical_2007,tailleur_simulation_2009}.
Here, the cloning is performed at each change of state of a copy.
The time intervals $\Delta t$ which separate those changes of state are non-infinitesimal, which
means that the formulation we used in~\cite{partI} cannot be applied to understand its convergence. 
Furthermore, because these time intervals are of non-constant duration and stochastically distributed, the continuous-time algorithm is more difficult to handle analytically than the discrete-time version. 
Instead of pursuing the analytical study within the continuous-time algorithm, we perform a numerical study, and we show that the $1/N_c$ and $1/t$ scalings are also observed for the continuous-time algorithm. 
%
%
%
Although the proof of these scalings are beyond the scope of the current paper, 
these numerical observations support a conjecture that such scaling in large $t$ and in large $N_c$ limits are generally valid in cloning algorithms to calculate large deviation functions.

\section{A Different CGF Estimator
\label{sec:two-estimators}
}

Normally, CGF estimator is defined as an arithmetic mean over many realizations, as seen in (\ref{eq:PSI1}). 
Here we show that another definition of the CGF estimator can be used, which
indeed provides better results than the ones from the standard estimator (in some parameter ranges). We define a new estimator as
\begin{equation} \label{eq:Lav}
 \Phi_{s}^{(N_{c})}  = \frac{1}{T} \log \overline{ \prod \limits_{i = 1}^{K_{r}} X_{i}^{r} },
\end{equation}
%
where we note that the average with respect to realizations are taken {\it inside} the logarithm.
As we discussed in Sec.~IV.C of~\cite{partI}, this estimator provides a correct value of CGF $\psi(s)$ in the infinite-time infinite-$N_c$ limits. This is thanks to the fact that the distribution of $\Psi_{s}^{(N_{c})}$ concentrates around $\psi(s)$ in those limits (the so-called ``self-averaging'' property).
At any finite population, one can rewrite $\Phi_{s}^{(N_{c})} $ using the large-time LDF principle~\eqref{eq:LDP1} as follows:
\begin{align} \label{eq:Lav2}
 \Phi_{s}^{(N_{c})} 
& = \frac{1}{T} \log \overline{ e^{T \Psi_{s}^{(N_{c})}}}
\\
& = \frac{1}{T} \log 
\int d\Psi
\
e^{-T \left[I_{N_c}(\Psi)+\Psi\vphantom{|^I}\right]}
\end{align}
which proves that in the large-$T$ limit, 
\begin{equation} \label{eq:Lav3}
 \Phi_{s}^{(N_{c})} = \min_{\Psi} \left[I_{N_c}(\Psi)+\Psi\vphantom{|^I}\right],
\end{equation}
to be compared to
\begin{equation} \label{eq:Lav4}
\overline{ \Psi_{s}^{(N_{c})} } = \underset{\Psi}{\operatorname{argmin}}\ I_{N_c}(\Psi) .
\end{equation}

\begin{figure}[t]
\includegraphics[width=0.48\textwidth]{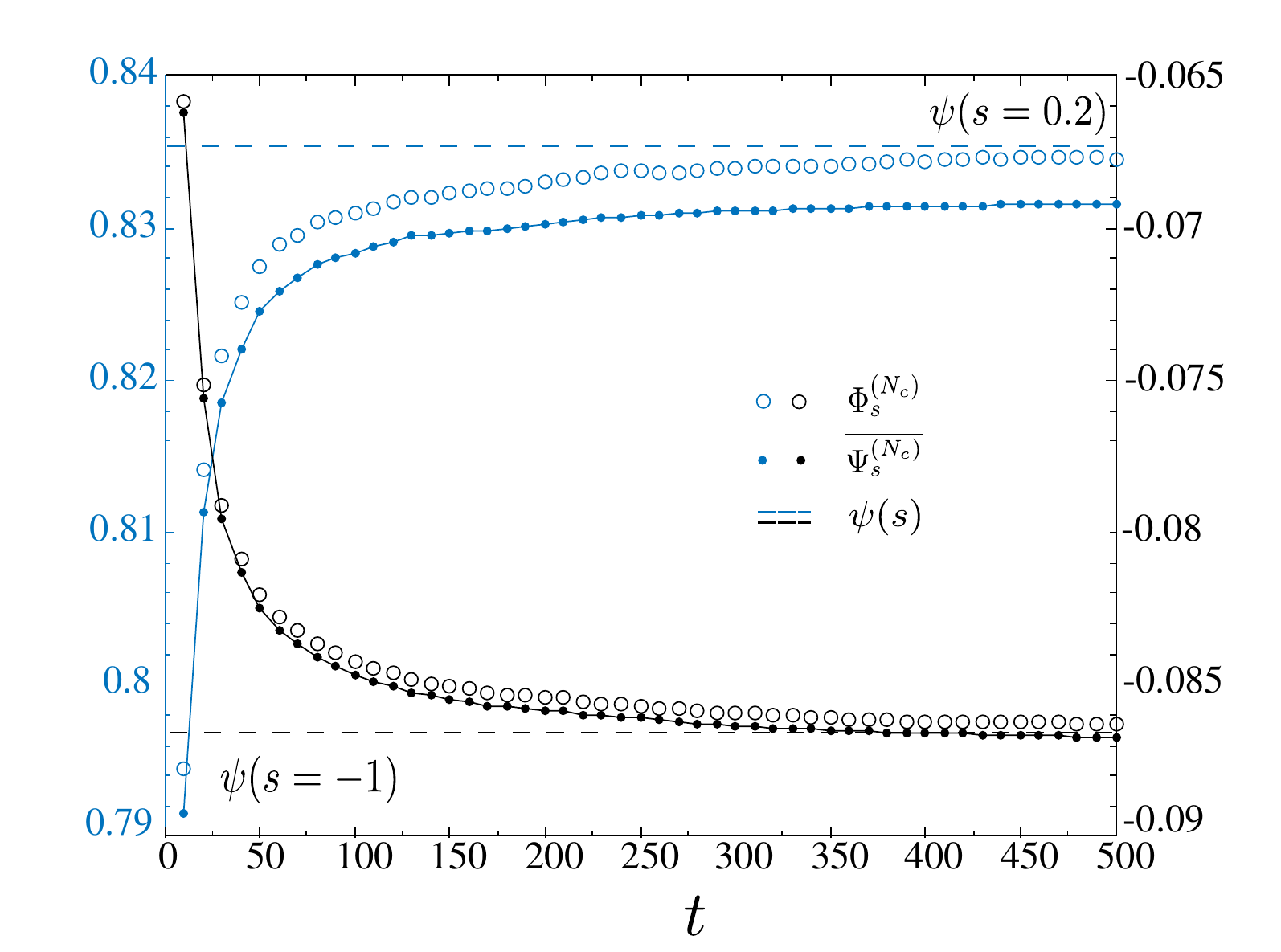}
\caption{\label{fig:avLLav} Comparison between two different estimators of the large deviation function, $\overline{ \Psi_{s}^{(N_{c})} }$ (equation~(\ref{eq:PSI1})) shown in dots and  $\Phi_{s}^{(N_{c})}$ (equation~(\ref{eq:Lav})) in circles, for the annihilation-creation dynamics~(\ref{eq:W_AC}). The analytical value  $\psi({s})$ (equation~(\ref{eq:PSIA})) is shown with a dashed line. Here we have also compared two different values of parameter $s=0.2$ (blue) and $s=-1$ (black). Additionally, $N_{c}=100$, $c=0.4$, $T = 500$ and $R=500$.
As discussed in the text, $\Phi_{s}^{(N_{c})}$ provides a better numerical evaluation of the CGF at small $s$.
}
\end{figure}

\begin{figure*}[t]
\includegraphics[width=0.32\textwidth]{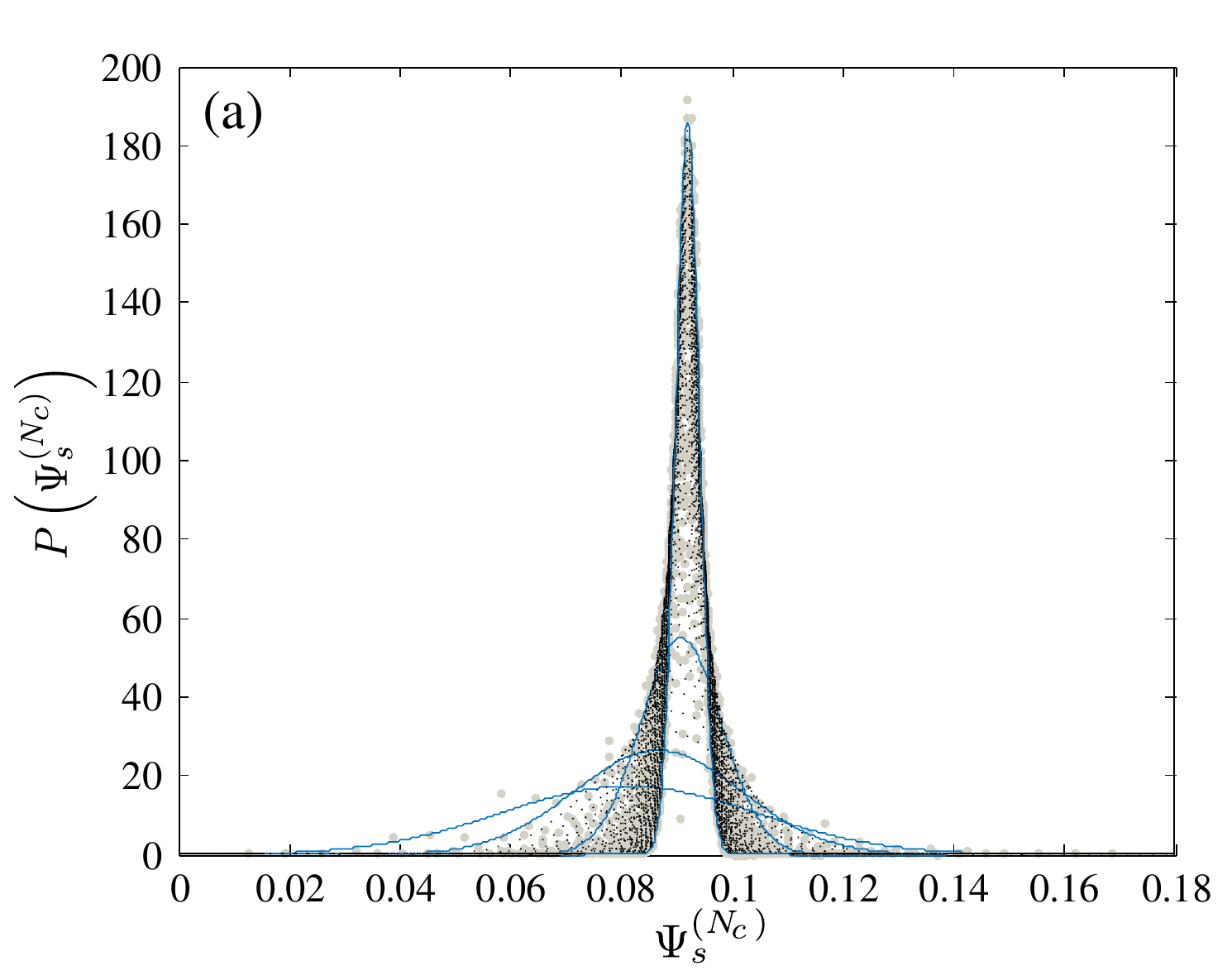}
\includegraphics[width=0.32\textwidth]{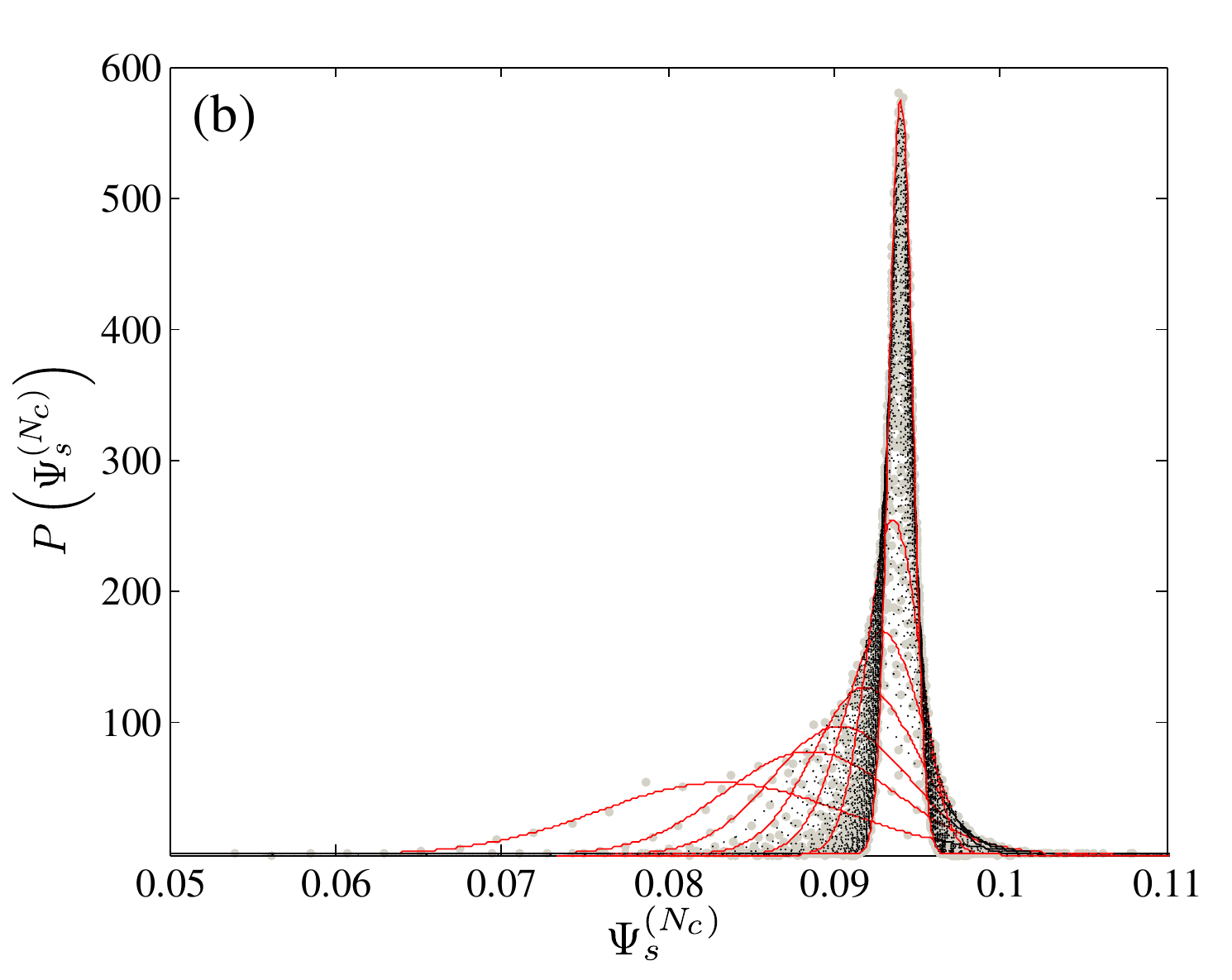}
\includegraphics[width=0.32\textwidth]{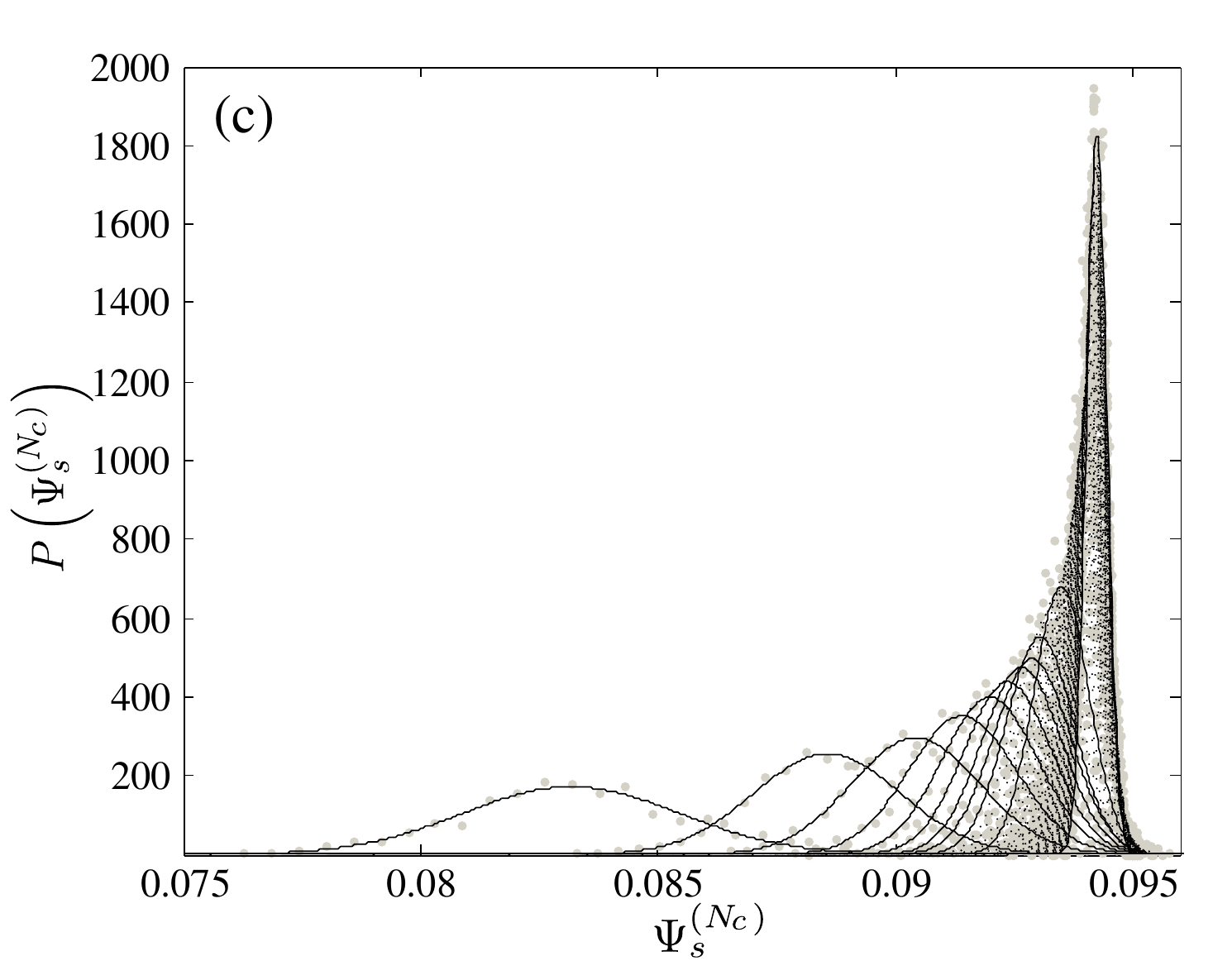}
\caption{\label{fig:PsiDistribution2} Distribution $P \left(  \Psi_s^{(N_{c})}  \right)$ of the CGF estimator $\Psi_s^{(N_{c})}$ for \textbf{(a)} $N_{c}=10$, \textbf{(b)} $N_{c}=100$ and \textbf{(c)} $N_{c}=1000$ and for simulation times $t \in \left[10,1000 \right]$. Each realization ($R=10^{4}$ for each simulation time) is shown with gray dots meanwhile its respective Gaussian fit (equation~(\ref{eq:DPSI1})) is shown with a dotted or a continuous curve. The dispersion of  $\Psi_s^{(N_{c})}$ is wider for shorter simulation times and small $N_{c}$. The mean value of the distribution converges to the theoretical value as the simulation time and the number of clones increase.}
\end{figure*}

On one hand, the definition~\eqref{eq:Lav} amounts to estimate $\psi$ from the exponential growth rate of the average of the final-$T$ population of many small (non-interacting) ``\mbox{islands}'', where the cloning algorithm would be operated. 
On the other hand, the estimator~\eqref{eq:PSI1} amounts to estimate $\psi$ from growth rate of a large ``island'' gathering the full set of the $R$ populations. The later is thus expected to be a better estimator of $\psi(s)$ than the former because it corresponds to a large population, where finite-size effects are less important. As a consequence, the estimator $\Phi_{s}^{(N_{c})}$ appears \emph{a priori} to be worse estimator than $\overline{ \Psi_{s}^{(N_{c})} }$ of $\psi(s)$.
However, as shown in Sec.~IV.C of~\cite{partI}, at small~$|s|$ and finite-$N_c$, a supplementary bias introduced by taking~\eqref{eq:Lav} in fact \emph{compensates} the finite-$N_c$ systematic error presented by~\eqref{eq:PSI1}, for a simple two state model.
Namely, the error is $O(sN_c^{-1})$ for~\eqref{eq:PSI1} while it is $O(s^2N_c^{-1})$ for~\eqref{eq:Lav}.
This fact is illustrated on Fig.~\ref{fig:avLLav}, where we show that at small $s=0.2$, $\Phi_{s}^{(N_{c})}$ provides a better estimation of $\psi(s)$ than $\overline{ \Psi_{s}^{(N_{c})} }$, while at larger $|s|$ ($s=-1$) the two estimators yield a comparable error.

\section{Fluctuations of CGF Estimator \label{sec:PSIDt}}
\subsection{Central Limit Theorem}
From relation~(\ref{eq:PSI1}), one can infer that the dispersion of the distribution of $\Psi_{s}^{(N_{c})}$ depends on the simulation time~$t$. This determines whether or not a large number of realizations $R$ is required in order to minimize the statistical error. In fact, as seen in Fig.~\ref{fig:PsiDistribution2}, the dispersion of $\Psi_{s}^{(N_{c})}$  is concentrated around its mean value, which approaches the analytical value $\psi(s)$ as the simulation time and the number of clones increase. 
\begin{figure}[b]
\includegraphics[width=0.48\textwidth]{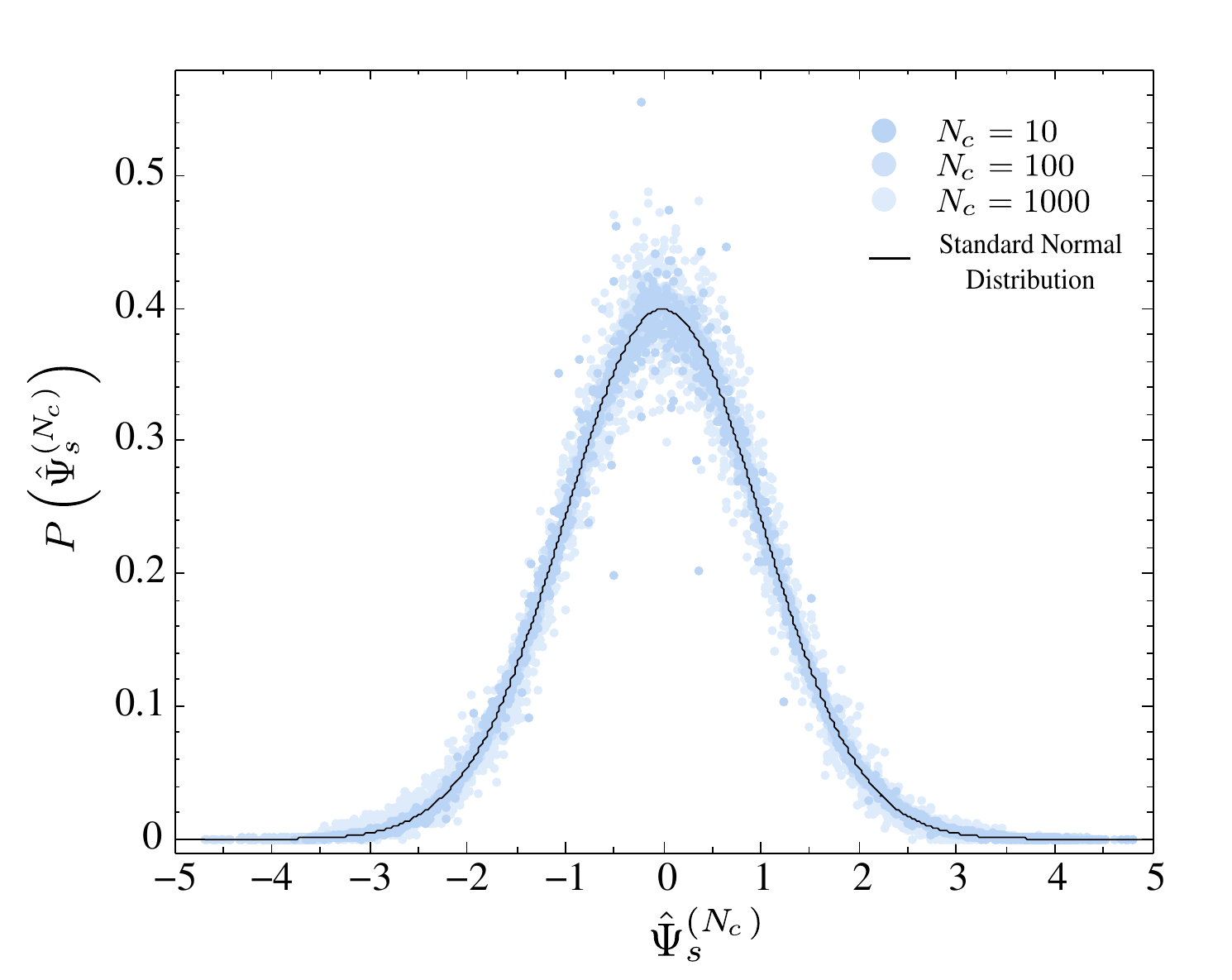}
\caption{\label{fig:NormDistribution} The distribution function of the rescaled variable $\hat{\Psi}_s^{(N_{c})}$ (equation~(\ref{eq:PSInorm})). Compatible with the central limit theorem, a collapse of the distribution function into a standard normal distribution for different number of clones is observed.}
\end{figure}

\begin{figure*}[t]
\includegraphics[width=0.48\textwidth]{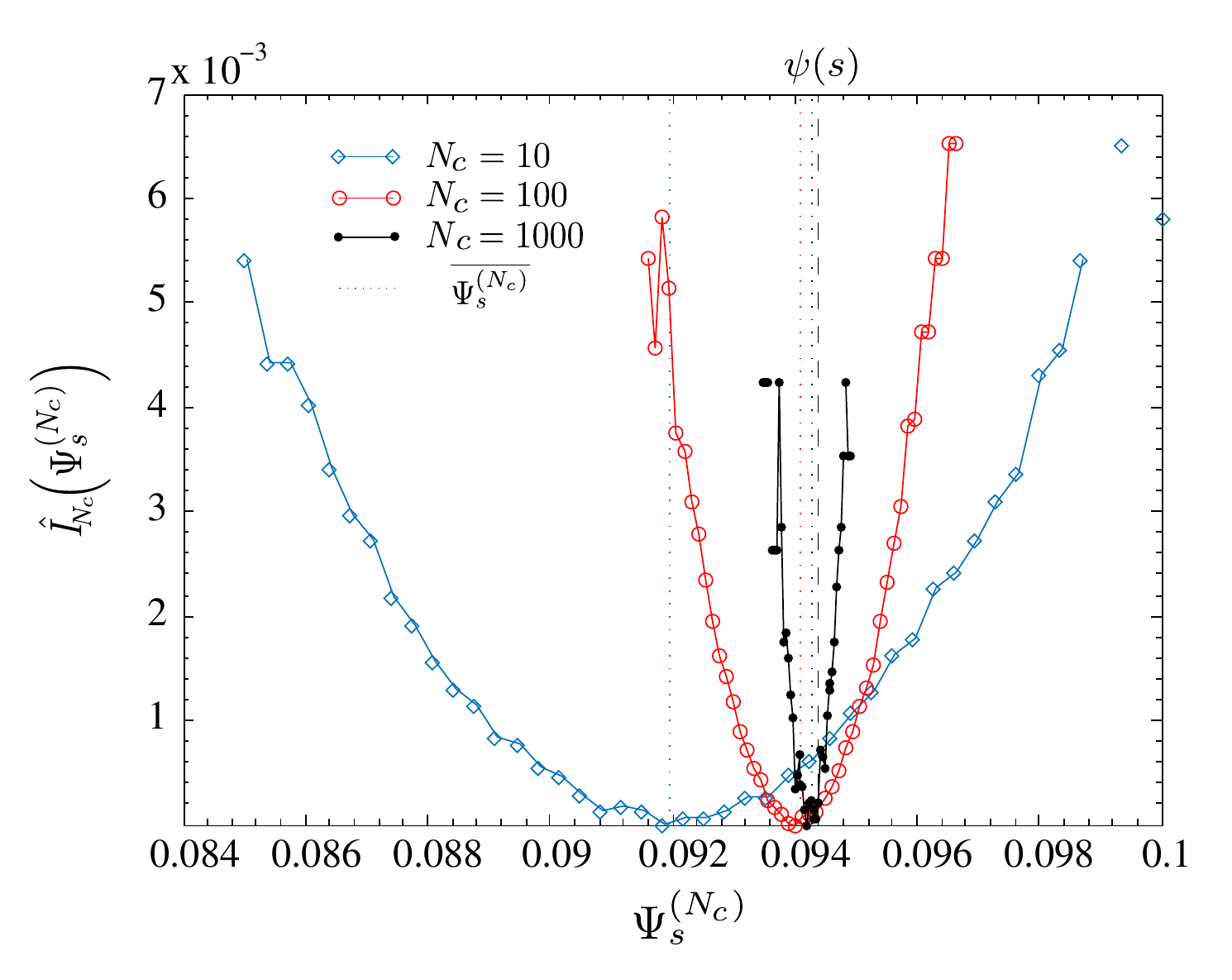}
\includegraphics[width=0.48\textwidth]{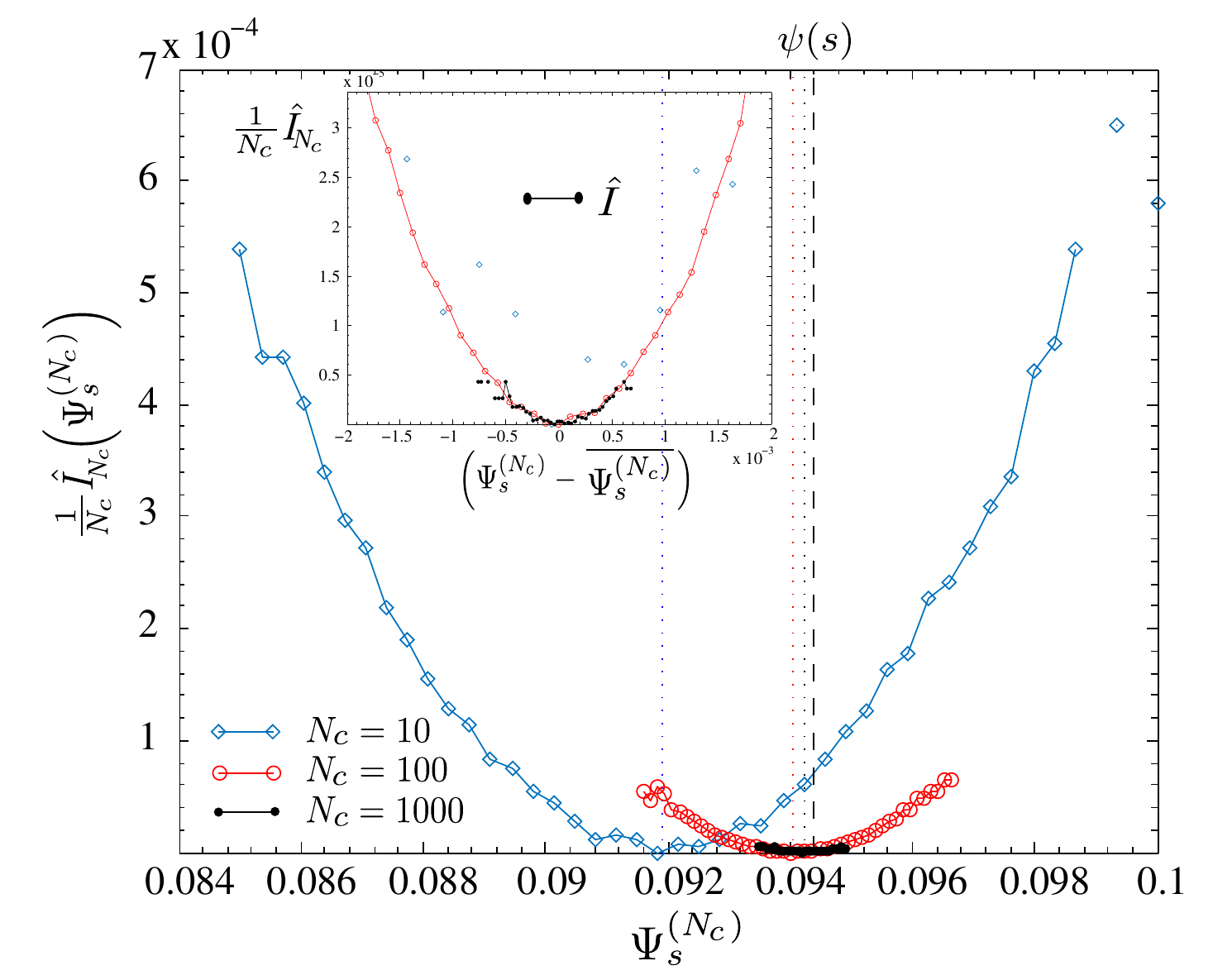}
\caption{\label{fig:I1} \textbf{(a)} Logarithmic distribution $\hat{I}_{N_c} \big( \Psi_s^{(N_{c})} \big)$  (equation~(\ref{eq:Ihat})).
Numerical evaluations were made for three fixed population sizes $N_{c} \in \{10,100,1000 \}$ with a fixed simulation time $T=1000$.
The logarithmic distribution presents a smaller width as $N_c$ increases.
The average over $R$ realizations of the CGF estimator $\overline{ \Psi_{s}^{(N_{c})} (T) }$ corresponds to the minimum of $\hat{I}_{N_c} \big( \Psi_s^{(N_{c})} \big)$ (dotted lines) and converges to the analytical value $\psi(s)$  (dashed lines) as $N_c\to\infty$. 
\textbf{(b)} Rescaled logarithmic distribution $\frac{1}{N_{c}}\hat{I}_{N_c} \big( \Psi_s^{(N_{c})} \big)$ as a function of $\Psi_s^{(N_{c})}$ and as a function of $\check{\Psi}_{s}^{(N_{c})} = \big( \Psi_s^{(N_{c})} - \overline{ \Psi_{s}^{(N_{c})}} \big)$ (inset) for a final simulation time $T=1000$.} 
\end{figure*}

We numerically confirm that these distributions are well-approximated by a Gaussian distribution
\begin{equation} \label{eq:DPSI1}
P \left(  \Psi_s^{(N_{c})}  \right) \sim A\, e^{-\frac{1}{C^{2}} \left( \Psi_s^{(N_{c})} -B \right) ^2}
\end{equation}
where the parameter $B$ is equal to $\overline{\Psi_{s}^{(N_{c})}(T)}$ and the parameters $A$ and $1/C^2$ are respectively of the order of $N_{c}^{1/2}$ and $N_{c}$. 
%
%
A mathematical argument to explain this obtained Gaussian distribution is given as follows: At any given time (not necessarily at $T$), let us perform the following rescaling
\begin{equation} \label{eq:PSInorm}
 \hat{\Psi}_s^{(N_{c})} = \frac{\Psi_{s}^{(N_{c})} - \overline{ \Psi_{s}^{(N_{c})} }}{\sigma_{\Psi_{s}^{(N_{c})}}},
\end{equation}
where $\sigma_{\Psi_{s}^{(N_{c})}}^{2}$
%
is the variance of the $R$ realizations of $\Psi_{s}^{(N_{c})}$.
Then, it produces a collapse of the distributions $P \big(  \hat{\Psi}_s^{(N_{c})} \big)$, for any  $t$ and any $N_{c}$ (Fig.~\ref{fig:NormDistribution}). We remark then that the CGF estimator (\ref{eq:PSI1}) is an additive observable of the history of the population, which follows a Markov dynamics. Hence, the rescaled estimator $\hat{\Psi}_s^{(N_{c})} $ follows a standard normal distribution in the large time limit, according to the central limit theorem (CLT):
\begin{equation} \label{eq:DPSI2}
P \big(  \hat{\Psi}_s^{(N_{c})} \big) = \frac{1}{\sqrt{2 \pi}} e^{-\frac{1}{2} \big( \hat{\Psi}_s^{(N_{c})} \big) ^2}.
\end{equation}

We note that this check of the CLT allows us to ensure if the steady-state of the population dynamics has been reached (note that in general the typical convergence time to the steady state is larger than the inverse of the spectral gap of the biased evolution operator \cite{hidalgo_discreteness_2016}).

By considering the scaling (\ref{eq:PSInorm}) we focus only on the small fluctuations of $\Psi_{s}^{(N_{c})}$ around $\overline{ \Psi_{s}^{(N_{c})} }$. But in general, the distribution function is not Gaussian, and in that case we need to consider a large deviation principle as below.

\subsection{Logarithmic Distribution of CGF Estimator}
Since $\Psi_{s}^{(N_{c})}$ is itself an additive observable of the dynamics of the ensemble of clones~\cite{partI}, the distribution of the CGF estimator $\Psi_{s}^{(N_{c})}$ satisfies itself a large deviation principle
\begin{equation} \label{eq:LDP1}
P \big(  \Psi_s^{(N_{c})}  \big) \sim e^{-t\: I_{N_c}\big( \Psi_s^{(N_{c})}  \big)},
\end{equation}
where $I_{N_c}\big( \Psi_s^{(N_{c})}  \big)$ is the rate function. This 
rate function could be evaluated in principle from the empirical distribution $P \big(\Psi_s^{(N_{c})} \big)$ as
\begin{equation} \label{eq:I1}
I_{N_c} \big( \Psi_s^{(N_{c})} \big) \approx -\frac{1}{t} \log P \big(  \Psi_s^{(N_{c})}  \big)  
\end{equation}
for a large $t$. 
Here we try to estimate the rate function from this equation. 
The numerical estimation of the right-hand side of the last expression at final simulation time $T$ is shown in Fig.~\ref{fig:I1}(a), where we have defined
\begin{equation} \label{eq:Ihat}
\hat{I}_{N_c} \big( \Psi_s^{(N_{c})} \big) \equiv -\frac{1}{t} \log P \big(  \Psi_s^{(N_{c})}  \big)   + \frac{1}{t} \log P \big(  \overline{ \Psi_{s}^{(N_{c})} }   \big)  
\end{equation}
so that $\hat{I}_{N_c} \big( \overline{ \Psi_{s}^{(N_{c})} } \big) = 0$. 
In the same figure, we also show $\overline{ \Psi_{s}^{(N_{c})} (T) }$ as vertical dotted lines which correspond to the minima of the logarithmic distribution $\hat{I}_{N_c} \big( \Psi_s^{(N_{c})} \big)$. As can be seen, these minima are displaced towards the analytical value $\psi(s)$ (shown with a dashed line) as \mbox{$N_{c}\to\infty$}. The logarithmic distribution $\hat I_{N_c}$ also becomes more concentrated as $N_{c}$ increases. 
%


Next, in order to study this decreasing of the width, 
we show
%
%
%
%
a rescaled logarithmic distribution
function $(1/N_c)\hat{I}_{N_c} \big( \Psi_s^{(N_{c})} \big)$
in Fig.~\ref{fig:I1}(b).
The minimum converges to the analytical value $\psi(s)$ (black dashed line) as $N_{c}\to\infty$. 
In the infinite-time infinite-size limit of $\Psi_s^{(N_{c})}$, it would be thus compatible with a logarithmic distribution function given by
\begin{equation} \label{eq:Int}
I \big( \Psi_s^{(N_{c})} \big) = - \lim_{N_c\to\infty} \:\frac{1}{N_c}\: \lim_{t\to\infty} \frac{1}{t} \log P \big(  \Psi_s^{(N_{c})} (t) \big) 
\end{equation}
which is shown (rescaled) with black dots in Fig.~\ref{fig:I1}(b). 
By performing the shift $\check{\Psi}_{s}^{(N_{c})} = \big( \Psi_s^{(N_{c})} - \overline{ \Psi_{s}^{(N_{c})} }  \big)$ we can see in the inset of Fig.~\ref{fig:I1}(b) the superposition of quadratic deviations of the numerical estimator $\Psi_{s}^{(N_{c})}$ around the minimum of $\hat{I_{N_c}}$ (especially for $N_c=100, 1000$). This indicates the decreasing of the fluctuation of CGF estimator proportional with both of $T$ and $N_c$ (see \cite{partI} for more detailed explanation). 
%


The obtained logarithmic distribution is well-approximated by a quadratic form, although these large deviations are in general not quadratic~\cite{partI}. 
This means that the direct observation discussed here cannot capture the large deviations of the CGF estimator (see also~\cite{rohwer_convergence_2015} for more detailed study of the direct estimation of rate functions).  
However we note that, for practical usage of the algorithm, we only consider small fluctuations described by central limit theorem,  although these large fluctuations might play an important role in more complicated systems, such as the ones presenting dynamical phase transitions.

\bibliography{ARTICLE.bib}

\begin{thebibliography}{24}%
\makeatletter
\providecommand \@ifxundefined [1]{%
 \@ifx{#1\undefined}
}%
\providecommand \@ifnum [1]{%
 \ifnum #1\expandafter \@firstoftwo
 \else \expandafter \@secondoftwo
 \fi
}%
\providecommand \@ifx [1]{%
 \ifx #1\expandafter \@firstoftwo
 \else \expandafter \@secondoftwo
 \fi
}%
\providecommand \natexlab [1]{#1}%
\providecommand \enquote  [1]{``#1''}%
\providecommand \bibnamefont  [1]{#1}%
\providecommand \bibfnamefont [1]{#1}%
\providecommand \citenamefont [1]{#1}%
\providecommand \href@noop [0]{\@secondoftwo}%
\providecommand \href [0]{\begingroup \@sanitize@url \@href}%
\providecommand \@href[1]{\@@startlink{#1}\@@href}%
\providecommand \@@href[1]{\endgroup#1\@@endlink}%
\providecommand \@sanitize@url [0]{\catcode `\\12\catcode `\$12\catcode
  `\&12\catcode `\#12\catcode `\^12\catcode `\_12\catcode `\%12\relax}%
\providecommand \@@startlink[1]{}%
\providecommand \@@endlink[0]{}%
\providecommand \url  [0]{\begingroup\@sanitize@url \@url }%
\providecommand \@url [1]{\endgroup\@href {#1}{\urlprefix }}%
\providecommand \urlprefix  [0]{URL }%
\providecommand \Eprint [0]{\href }%
\providecommand \doibase [0]{http://dx.doi.org/}%
\providecommand \selectlanguage [0]{\@gobble}%
\providecommand \bibinfo  [0]{\@secondoftwo}%
\providecommand \bibfield  [0]{\@secondoftwo}%
\providecommand \translation [1]{[#1]}%
\providecommand \BibitemOpen [0]{}%
\providecommand \bibitemStop [0]{}%
\providecommand \bibitemNoStop [0]{.\EOS\space}%
\providecommand \EOS [0]{\spacefactor3000\relax}%
\providecommand \BibitemShut  [1]{\csname bibitem#1\endcsname}%
\let\auto@bib@innerbib\@empty
\bibitem [{\citenamefont {Kahn}\ and\ \citenamefont
  {Harris}(1951)}]{kahn1951estimation}%
  \BibitemOpen
  \bibfield  {author} {\bibinfo {author} {\bibfnamefont {H.}~\bibnamefont
  {Kahn}}\ and\ \bibinfo {author} {\bibfnamefont {T.~E.}\ \bibnamefont
  {Harris}},\ }\href@noop {} {\bibfield  {journal} {\bibinfo  {journal}
  {National Bureau of Standards applied mathematics series}\ }\textbf {\bibinfo
  {volume} {12}},\ \bibinfo {pages} {27} (\bibinfo {year} {1951})}\BibitemShut
  {NoStop}%
\bibitem [{\citenamefont {C\'erou}\ and\ \citenamefont
  {Guyader}(2007)}]{cerou_adaptive_2007}%
  \BibitemOpen
  \bibfield  {author} {\bibinfo {author} {\bibfnamefont {F.}~\bibnamefont
  {C\'erou}}\ and\ \bibinfo {author} {\bibfnamefont {A.}~\bibnamefont
  {Guyader}},\ }\href {\doibase 10.1080/07362990601139628} {\bibfield
  {journal} {\bibinfo  {journal} {Stochastic Analysis and Applications}\
  }\textbf {\bibinfo {volume} {25}},\ \bibinfo {pages} {417} (\bibinfo {year}
  {2007})}\BibitemShut {NoStop}%
\bibitem [{\citenamefont {Bolhuis}\ \emph {et~al.}(2002)\citenamefont
  {Bolhuis}, \citenamefont {Chandler}, \citenamefont {Dellago},\ and\
  \citenamefont {Geissler}}]{bolhuis_transition_2002}%
  \BibitemOpen
  \bibfield  {author} {\bibinfo {author} {\bibfnamefont {P.~G.}\ \bibnamefont
  {Bolhuis}}, \bibinfo {author} {\bibfnamefont {D.}~\bibnamefont {Chandler}},
  \bibinfo {author} {\bibfnamefont {C.}~\bibnamefont {Dellago}}, \ and\
  \bibinfo {author} {\bibfnamefont {P.~L.}\ \bibnamefont {Geissler}},\ }\href
  {\doibase 10.1146/annurev.physchem.53.082301.113146} {\bibfield  {journal}
  {\bibinfo  {journal} {Annual Review of Physical Chemistry}\ }\textbf
  {\bibinfo {volume} {53}},\ \bibinfo {pages} {291} (\bibinfo {year}
  {2002})}\BibitemShut {NoStop}%
\bibitem [{\citenamefont {Bucklew}(2013)}]{bucklew_introduction_2013}%
  \BibitemOpen
  \bibfield  {author} {\bibinfo {author} {\bibfnamefont {J.}~\bibnamefont
  {Bucklew}},\ }\href@noop {} {\emph {\bibinfo {title} {Introduction to {Rare}
  {Event} {Simulation}}}}\ (\bibinfo  {publisher} {Springer Science \& Business
  Media},\ \bibinfo {year} {2013})\BibitemShut {NoStop}%
\bibitem [{\citenamefont {Giardin\`a}\ \emph {et~al.}(2011)\citenamefont
  {Giardin\`a}, \citenamefont {Kurchan}, \citenamefont {Lecomte},\ and\
  \citenamefont {Tailleur}}]{giardina_simulating_2011}%
  \BibitemOpen
  \bibfield  {author} {\bibinfo {author} {\bibfnamefont {C.}~\bibnamefont
  {Giardin\`a}}, \bibinfo {author} {\bibfnamefont {J.}~\bibnamefont {Kurchan}},
  \bibinfo {author} {\bibfnamefont {V.}~\bibnamefont {Lecomte}}, \ and\
  \bibinfo {author} {\bibfnamefont {J.}~\bibnamefont {Tailleur}},\ }\href
  {\doibase 10.1007/s10955-011-0350-4} {\bibfield  {journal} {\bibinfo
  {journal} {J. Stat. Phys.}\ }\textbf {\bibinfo {volume} {145}},\ \bibinfo
  {pages} {787} (\bibinfo {year} {2011})}\BibitemShut {NoStop}%
\bibitem [{\citenamefont {Giardin{\`a}}\ \emph {et~al.}(2006)\citenamefont
  {Giardin{\`a}}, \citenamefont {Kurchan},\ and\ \citenamefont
  {Peliti}}]{giardina_direct_2006}%
  \BibitemOpen
  \bibfield  {author} {\bibinfo {author} {\bibfnamefont {C.}~\bibnamefont
  {Giardin{\`a}}}, \bibinfo {author} {\bibfnamefont {J.}~\bibnamefont
  {Kurchan}}, \ and\ \bibinfo {author} {\bibfnamefont {L.}~\bibnamefont
  {Peliti}},\ }\href {\doibase 10.1103/PhysRevLett.96.120603} {\bibfield
  {journal} {\bibinfo  {journal} {Phys. Rev. Lett.}\ }\textbf {\bibinfo
  {volume} {96}},\ \bibinfo {pages} {120603} (\bibinfo {year}
  {2006})}\BibitemShut {NoStop}%
\bibitem [{\citenamefont {Tailleur}\ and\ \citenamefont
  {Kurchan}(2007)}]{tailleur_probing_2007}%
  \BibitemOpen
  \bibfield  {author} {\bibinfo {author} {\bibfnamefont {J.}~\bibnamefont
  {Tailleur}}\ and\ \bibinfo {author} {\bibfnamefont {J.}~\bibnamefont
  {Kurchan}},\ }\href {\doibase 10.1038/nphys515} {\bibfield  {journal}
  {\bibinfo  {journal} {Nat Phys}\ }\textbf {\bibinfo {volume} {3}},\ \bibinfo
  {pages} {203} (\bibinfo {year} {2007})}\BibitemShut {NoStop}%
\bibitem [{\citenamefont {{Guevara~Hidalgo}}\ and\ \citenamefont
  {Lecomte}(2016)}]{hidalgo_discreteness_2016}%
  \BibitemOpen
  \bibfield  {author} {\bibinfo {author} {\bibfnamefont {E.}~\bibnamefont
  {{Guevara~Hidalgo}}}\ and\ \bibinfo {author} {\bibfnamefont {V.}~\bibnamefont
  {Lecomte}},\ }\href {\doibase 10.1088/1751-8113/49/20/205002} {\bibfield
  {journal} {\bibinfo  {journal} {J. Phys. A: Math. Theor.}\ }\textbf {\bibinfo
  {volume} {49}},\ \bibinfo {pages} {205002} (\bibinfo {year}
  {2016})}\BibitemShut {NoStop}%
\bibitem [{\citenamefont {Hurtado}\ and\ \citenamefont
  {Garrido}(2009)}]{hurtado_current_2009}%
  \BibitemOpen
  \bibfield  {author} {\bibinfo {author} {\bibfnamefont {P.~I.}\ \bibnamefont
  {Hurtado}}\ and\ \bibinfo {author} {\bibfnamefont {P.~L.}\ \bibnamefont
  {Garrido}},\ }\href {\doibase 10.1088/1742-5468/2009/02/P02032} {\bibfield
  {journal} {\bibinfo  {journal} {J. Stat. Mech.}\ }\textbf {\bibinfo {volume}
  {2009}},\ \bibinfo {pages} {P02032} (\bibinfo {year} {2009})}\BibitemShut
  {NoStop}%
\bibitem [{\citenamefont {Tchernookov}\ and\ \citenamefont
  {Dinner}(2010)}]{tchernookov_list-based_2010}%
  \BibitemOpen
  \bibfield  {author} {\bibinfo {author} {\bibfnamefont {M.}~\bibnamefont
  {Tchernookov}}\ and\ \bibinfo {author} {\bibfnamefont {A.~R.}\ \bibnamefont
  {Dinner}},\ }\href {\doibase 10.1088/1742-5468/2010/02/P02006} {\bibfield
  {journal} {\bibinfo  {journal} {J. Stat. Mech.}\ }\textbf {\bibinfo {volume}
  {2010}},\ \bibinfo {pages} {P02006} (\bibinfo {year} {2010})}\BibitemShut
  {NoStop}%
\bibitem [{\citenamefont {Kundu}\ \emph {et~al.}(2011)\citenamefont {Kundu},
  \citenamefont {Sabhapandit},\ and\ \citenamefont
  {Dhar}}]{kundu_application_2011}%
  \BibitemOpen
  \bibfield  {author} {\bibinfo {author} {\bibfnamefont {A.}~\bibnamefont
  {Kundu}}, \bibinfo {author} {\bibfnamefont {S.}~\bibnamefont {Sabhapandit}},
  \ and\ \bibinfo {author} {\bibfnamefont {A.}~\bibnamefont {Dhar}},\ }\href
  {\doibase 10.1103/PhysRevE.83.031119} {\bibfield  {journal} {\bibinfo
  {journal} {Phys. Rev. E}\ }\textbf {\bibinfo {volume} {83}},\ \bibinfo
  {pages} {031119} (\bibinfo {year} {2011})}\BibitemShut {NoStop}%
\bibitem [{\citenamefont {Nemoto}\ \emph {et~al.}(2016)\citenamefont {Nemoto},
  \citenamefont {Bouchet}, \citenamefont {Jack},\ and\ \citenamefont
  {Lecomte}}]{nemoto_population-dynamics_2016}%
  \BibitemOpen
  \bibfield  {author} {\bibinfo {author} {\bibfnamefont {T.}~\bibnamefont
  {Nemoto}}, \bibinfo {author} {\bibfnamefont {F.}~\bibnamefont {Bouchet}},
  \bibinfo {author} {\bibfnamefont {R.~L.}\ \bibnamefont {Jack}}, \ and\
  \bibinfo {author} {\bibfnamefont {V.}~\bibnamefont {Lecomte}},\ }\href
  {\doibase 10.1103/PhysRevE.93.062123} {\bibfield  {journal} {\bibinfo
  {journal} {Phys. Rev. E}\ }\textbf {\bibinfo {volume} {93}},\ \bibinfo
  {pages} {062123} (\bibinfo {year} {2016})}\BibitemShut {NoStop}%
\bibitem [{\citenamefont {Nemoto}\ \emph
  {et~al.}(2017{\natexlab{a}})\citenamefont {Nemoto}, \citenamefont
  {Guevara~Hidalgo},\ and\ \citenamefont {Lecomte}}]{partI}%
  \BibitemOpen
  \bibfield  {author} {\bibinfo {author} {\bibfnamefont {T.}~\bibnamefont
  {Nemoto}}, \bibinfo {author} {\bibfnamefont {E.}~\bibnamefont
  {Guevara~Hidalgo}}, \ and\ \bibinfo {author} {\bibfnamefont {V.}~\bibnamefont
  {Lecomte}},\ }\href {\doibase 10.1103/PhysRevE.95.012102} {\bibfield
  {journal} {\bibinfo  {journal} {Phys. Rev. E}\ }\textbf {\bibinfo {volume}
  {95}},\ \bibinfo {pages} {012102} (\bibinfo {year}
  {2017}{\natexlab{a}})}\BibitemShut {NoStop}%
\bibitem [{\citenamefont {Lecomte}\ and\ \citenamefont
  {Tailleur}(2007)}]{lecomte_numerical_2007}%
  \BibitemOpen
  \bibfield  {author} {\bibinfo {author} {\bibfnamefont {V.}~\bibnamefont
  {Lecomte}}\ and\ \bibinfo {author} {\bibfnamefont {J.}~\bibnamefont
  {Tailleur}},\ }\href {\doibase 10.1088/1742-5468/2007/03/P03004} {\bibfield
  {journal} {\bibinfo  {journal} {J. Stat. Mech.}\ }\textbf {\bibinfo {volume}
  {2007}},\ \bibinfo {pages} {P03004} (\bibinfo {year} {2007})}\BibitemShut
  {NoStop}%
\bibitem [{\citenamefont {Tailleur}\ and\ \citenamefont
  {Lecomte}(2009)}]{tailleur_simulation_2009}%
  \BibitemOpen
  \bibfield  {author} {\bibinfo {author} {\bibfnamefont {J.}~\bibnamefont
  {Tailleur}}\ and\ \bibinfo {author} {\bibfnamefont {V.}~\bibnamefont
  {Lecomte}},\ }\href {\doibase 10.1063/1.3082284} {\bibfield  {journal}
  {\bibinfo  {journal} {AIP Conf. Proc.}\ }\textbf {\bibinfo {volume} {1091}},\
  \bibinfo {pages} {212} (\bibinfo {year} {2009})}\BibitemShut {NoStop}%
\bibitem [{\citenamefont {Touchette}(2009)}]{touchette_large_2009}%
  \BibitemOpen
  \bibfield  {author} {\bibinfo {author} {\bibfnamefont {H.}~\bibnamefont
  {Touchette}},\ }\href {\doibase 10.1016/j.physrep.2009.05.002} {\bibfield
  {journal} {\bibinfo  {journal} {Physics Reports}\ }\textbf {\bibinfo {volume}
  {478}},\ \bibinfo {pages} {1} (\bibinfo {year} {2009})}\BibitemShut {NoStop}%
\bibitem [{\citenamefont {Garrahan}\ \emph {et~al.}(2009)\citenamefont
  {Garrahan}, \citenamefont {Jack}, \citenamefont {Lecomte}, \citenamefont
  {Pitard}, \citenamefont {van Duijvendijk},\ and\ \citenamefont {van
  Wijland}}]{garrahan_first-order_2009}%
  \BibitemOpen
  \bibfield  {author} {\bibinfo {author} {\bibfnamefont {J.~P.}\ \bibnamefont
  {Garrahan}}, \bibinfo {author} {\bibfnamefont {R.~L.}\ \bibnamefont {Jack}},
  \bibinfo {author} {\bibfnamefont {V.}~\bibnamefont {Lecomte}}, \bibinfo
  {author} {\bibfnamefont {E.}~\bibnamefont {Pitard}}, \bibinfo {author}
  {\bibfnamefont {K.}~\bibnamefont {van Duijvendijk}}, \ and\ \bibinfo {author}
  {\bibfnamefont {F.}~\bibnamefont {van Wijland}},\ }\href {\doibase
  10.1088/1751-8113/42/7/075007} {\bibfield  {journal} {\bibinfo  {journal} {J.
  Phys. A}\ }\textbf {\bibinfo {volume} {42}},\ \bibinfo {pages} {075007}
  (\bibinfo {year} {2009})}\BibitemShut {NoStop}%
\bibitem [{\citenamefont {Harris}(1974)}]{CP}%
  \BibitemOpen
  \bibfield  {author} {\bibinfo {author} {\bibfnamefont {T.~E.}\ \bibnamefont
  {Harris}},\ }\href {http://www.jstor.org/stable/2959099} {\bibfield
  {journal} {\bibinfo  {journal} {Ann. Probability}\ }\textbf {\bibinfo
  {volume} {2}},\ \bibinfo {pages} {969} (\bibinfo {year} {1974})}\BibitemShut
  {NoStop}%
\bibitem [{\citenamefont {Bezuidenhout}\ and\ \citenamefont
  {Grimmett}(1990)}]{10.2307/2244329}%
  \BibitemOpen
  \bibfield  {author} {\bibinfo {author} {\bibfnamefont {C.}~\bibnamefont
  {Bezuidenhout}}\ and\ \bibinfo {author} {\bibfnamefont {G.}~\bibnamefont
  {Grimmett}},\ }\href {http://www.jstor.org/stable/2244329} {\bibfield
  {journal} {\bibinfo  {journal} {The Annals of Probability}\ }\textbf
  {\bibinfo {volume} {18}},\ \bibinfo {pages} {1462} (\bibinfo {year}
  {1990})}\BibitemShut {NoStop}%
\bibitem [{\citenamefont {Lecomte}\ \emph {et~al.}(2007)\citenamefont
  {Lecomte}, \citenamefont {Appert-Rolland},\ and\ \citenamefont {van
  Wijland}}]{Lecomte2007}%
  \BibitemOpen
  \bibfield  {author} {\bibinfo {author} {\bibfnamefont {V.}~\bibnamefont
  {Lecomte}}, \bibinfo {author} {\bibfnamefont {C.}~\bibnamefont
  {Appert-Rolland}}, \ and\ \bibinfo {author} {\bibfnamefont {F.}~\bibnamefont
  {van Wijland}},\ }\href {\doibase 10.1007/s10955-006-9254-0} {\bibfield
  {journal} {\bibinfo  {journal} {Journal of Statistical Physics}\ }\textbf
  {\bibinfo {volume} {127}},\ \bibinfo {pages} {51} (\bibinfo {year}
  {2007})}\BibitemShut {NoStop}%
\bibitem [{\citenamefont {Rohwer}\ \emph {et~al.}(2015)\citenamefont {Rohwer},
  \citenamefont {Angeletti},\ and\ \citenamefont
  {Touchette}}]{rohwer_convergence_2015}%
  \BibitemOpen
  \bibfield  {author} {\bibinfo {author} {\bibfnamefont {C.~M.}\ \bibnamefont
  {Rohwer}}, \bibinfo {author} {\bibfnamefont {F.}~\bibnamefont {Angeletti}}, \
  and\ \bibinfo {author} {\bibfnamefont {H.}~\bibnamefont {Touchette}},\ }\href
  {\doibase 10.1103/PhysRevE.92.052104} {\bibfield  {journal} {\bibinfo
  {journal} {Phys. Rev. E}\ }\textbf {\bibinfo {volume} {92}},\ \bibinfo
  {pages} {052104} (\bibinfo {year} {2015})}\BibitemShut {NoStop}%
\bibitem [{\citenamefont {Kolton}\ \emph {et~al.}(2013)\citenamefont {Kolton},
  \citenamefont {Bustingorry}, \citenamefont {Ferrero},\ and\ \citenamefont
  {Rosso}}]{kolton_uniqueness_2013}%
  \BibitemOpen
  \bibfield  {author} {\bibinfo {author} {\bibfnamefont {A.~B.}\ \bibnamefont
  {Kolton}}, \bibinfo {author} {\bibfnamefont {S.}~\bibnamefont {Bustingorry}},
  \bibinfo {author} {\bibfnamefont {E.~E.}\ \bibnamefont {Ferrero}}, \ and\
  \bibinfo {author} {\bibfnamefont {A.}~\bibnamefont {Rosso}},\ }\href
  {\doibase 10.1088/1742-5468/2013/12/P12004} {\bibfield  {journal} {\bibinfo
  {journal} {J. Stat. Mech.}\ }\textbf {\bibinfo {volume} {2013}},\ \bibinfo
  {pages} {P12004} (\bibinfo {year} {2013})}\BibitemShut {NoStop}%
\bibitem [{\citenamefont {Nyawo}\ and\ \citenamefont
  {Touchette}(2016)}]{0295-5075-116-5-50009}%
  \BibitemOpen
  \bibfield  {author} {\bibinfo {author} {\bibfnamefont {P.~T.}\ \bibnamefont
  {Nyawo}}\ and\ \bibinfo {author} {\bibfnamefont {H.}~\bibnamefont
  {Touchette}},\ }\href {http://stacks.iop.org/0295-5075/116/i=5/a=50009}
  {\bibfield  {journal} {\bibinfo  {journal} {EPL (Europhysics Letters)}\
  }\textbf {\bibinfo {volume} {116}},\ \bibinfo {pages} {50009} (\bibinfo
  {year} {2016})}\BibitemShut {NoStop}%
\bibitem [{\citenamefont {Nemoto}\ \emph
  {et~al.}(2017{\natexlab{b}})\citenamefont {Nemoto}, \citenamefont {Jack},\
  and\ \citenamefont {Lecomte}}]{PhysRevLett.118.115702}%
  \BibitemOpen
  \bibfield  {author} {\bibinfo {author} {\bibfnamefont {T.}~\bibnamefont
  {Nemoto}}, \bibinfo {author} {\bibfnamefont {R.~L.}\ \bibnamefont {Jack}}, \
  and\ \bibinfo {author} {\bibfnamefont {V.}~\bibnamefont {Lecomte}},\ }\href
  {\doibase 10.1103/PhysRevLett.118.115702} {\bibfield  {journal} {\bibinfo
  {journal} {Phys. Rev. Lett.}\ }\textbf {\bibinfo {volume} {118}},\ \bibinfo
  {pages} {115702} (\bibinfo {year} {2017}{\natexlab{b}})}\BibitemShut
  {NoStop}%
\end{thebibliography}%

\end{document}